\documentclass[11pt,a4paper]{elsarticle}
\usepackage[T1]{fontenc}

\usepackage{amsmath}
\usepackage{amssymb}
\usepackage{bm}
\usepackage{epsfig}
\usepackage{graphicx}
\usepackage[colorlinks=true]{hyperref}
\usepackage{listings}
\usepackage{ragged2e}
\usepackage{subcaption}
\usepackage{tikz}
\usetikzlibrary{arrows}
\usetikzlibrary{decorations.markings}
\usetikzlibrary{decorations.pathmorphing}
\usetikzlibrary{decorations.pathreplacing}
\usetikzlibrary{shapes}
\usetikzlibrary{trees}

\input{all.tikzdefs}
\tikzstyle{blank}=[fill=white, shape=circle, draw=white, inner sep=1.0pt]
\tikzstyle{dot}=[fill=black, shape=circle, draw=black, inner sep=1.0pt]

\tikzstyle{line}=[-, draw=black!80!white, line width=1pt, line cap=rect]
\tikzstyle{thick line}=[-, draw=black, line width=1.5pt, line cap=rect]
\tikzstyle{thin line}=[-, draw=black, line width=0.6pt, line cap=rect]
\tikzstyle{external edge}=[-, line width=1pt, line cap=rect, draw={rgb,255: red,102; green,102; blue,102}]
\tikzstyle{massless external edge}=[-, line width=0.8pt, densely dashed, line cap=rect, draw={rgb,255: red,102; green,102; blue,102}]
\tikzstyle{edge}=[-, draw={rgb,255: red,176; green,36; blue,39}, line width=1pt, preaction={{draw=white,line width=2pt}}, line cap=rect]
\tikzstyle{massless edge}=[-, draw={rgb,255: red,176; green,36; blue,39}, line width=0.8pt, densely dashed, line cap=rect]
\tikzstyle{dot1}=[-, postaction=decorate, decoration={markings,mark=at position .50 with {\node[style=dot]{};}}]
\tikzstyle{dot2}=[-, postaction=decorate, decoration={markings,mark=between positions 0.33 and 0.67 step 0.33 with {\node[style=dot]{};}}]
\tikzstyle{dot3}=[-, postaction=decorate, decoration={markings,mark=between positions 0.25 and 0.76 step 0.25 with {\node[style=dot]{};}}]
\tikzstyle{dot4}=[-, postaction=decorate, decoration={markings,mark=between positions 0.20 and 0.81 step 0.20 with {\node[style=dot]{};}}]

\newcommand*{\sfig}[1]{{\scalebox{0.75}{{\input{#1}}}}}

\definecolor{grey}{RGB}{171,171,171}
\definecolor{EmeraldGreen}{HTML}{1ea78d}
\definecolor{EnglishRed}{HTML}{b02427}

\AtBeginDocument{\hypersetup{urlcolor=EmeraldGreen,citecolor=EmeraldGreen,linkcolor=EnglishRed}}

\usepackage{booktabs}
\usepackage{array}
\newcolumntype{M}[1]{>{\centering\arraybackslash}m{#1}}	% to center values in the table

\usepackage[super]{nth}
\definecolor{light-gray}{gray}{0.97}

\usepackage[frozencache=true,cachedir=mintedcachedir,newfloat]{minted}

\usepackage[colorinlistoftodos, textsize=scriptsize]{todonotes}
\newenvironment{code}{\captionsetup{type=listing}}{}
\SetupFloatingEnvironment{listing}{name=Code Example}
\graphicspath{{plots/}}
\newcounter{bla}
\newenvironment{refnummer}{%
\list{[\arabic{bla}]}%
{\usecounter{bla}%
 \setlength{\itemindent}{0pt}%
 \setlength{\topsep}{0pt}%
 \setlength{\itemsep}{0pt}%
 \setlength{\labelsep}{2pt}%
 \setlength{\listparindent}{0pt}%
 \settowidth{\labelwidth}{[9]}%
 \setlength{\leftmargin}{\labelwidth}%
 \addtolength{\leftmargin}{\labelsep}%
 \setlength{\rightmargin}{0pt}}}
 {\endlist}
\def\be{\begin{equation}}
\def\ee{\end{equation}}
\def\bea{\begin{align}}
\def\eea{\end{align}}
\def\nn{\nonumber}

\newcommand{\pysecdec}{py{\textsc{SecDec}}}

\newcommand{\form}{{\textsc{Form}}}

\newcommand{\eps}{\varepsilon}

\newcommand{\qmc}{\textsc{Qmc}}

\newcommand{\eqn}[1]{Eq.~(\ref{#1})}

\begin{document}

\begin{frontmatter}

\title{%
\vspace*{-2cm}\hfill{\normalsize\texttt{KA-TP-19-2021, P3H-21-055, MPP-2021-131,\\ \hfill IPPP/21/20, ZU-TH 35/21, PSI-PR-21-17}}\\\vspace*{1cm}
Expansion by regions with pySecDec}

\author[a]{G.~Heinrich}
\author[b]{S.~Jahn}
\author[c]{S.~P.~Jones}
\author[d]{M.~Kerner}
\author[b,e]{F.~Langer}
\author[a]{V.~Magerya}
\author[b,f]{A.~P\~oldaru}
\author[g]{J.~Schlenk}
\author[a]{E.~Villa}

\address[a]{Institute for Theoretical Physics, Karlsruhe Institute of Technology (KIT), 76128 Karlsruhe, Germany}
\address[b]{Max Planck Institute for Physics, F\"ohringer Ring 6, 80805 M\"unchen, Germany}
\address[c]{Institute for Particle Physics Phenomenology, Durham University, Durham DH1 3LE, UK}
\address[d]{Physik-Institut, Universit{\"a}t Z{\"u}rich, Winterthurerstrasse 190, 8057 Z{\"u}rich, Switzerland}
\address[e]{Department of Engineering, University of Cambridge, Cambridge CB2 1PZ, UK}
\address[f]{Faculty of Physics, Ludwig-Maximilians-Universit{\"a}t M{\"u}nchen, 80539 M\"unchen, Germany}
\address[g]{Theory Group LTP, Paul Scherrer Institut, CH-5232 Villigen PSI, Switzerland}

\begin{abstract}
We discuss the technique of expansion by regions from a geometric perspective, and its implementation within \pysecdec{}, a toolbox for the evaluation of dimensionally regulated parameter integrals.
The program offers an automated way to perform asymptotic expansions and provides a new mechanism for efficiently evaluating amplitudes, as well as individual integrals.
The usage of the new features available within \pysecdec{} is illustrated with several examples. 
\end{abstract}

%\begin{flushleft}
%PACS: 12.38.Bx, %perturbative calculations
%02.60.Jh, 	%Numerical differentiation and integration 
%02.70.Wz 	% Symbolic computation (computer algebra)
%\end{flushleft}


\end{frontmatter}

\newpage

{\bf PROGRAM SUMMARY}
   
\begin{small}
\noindent
{\em Manuscript Title: } Expansion by regions with py\textsc{SecDec}   \\
{\em Authors: } G.~Heinrich, S.~Jahn, S.~P.~Jones, M.~Kerner,
F.~Langer, V.~Magerya, A.~P\~oldaru, J.~Schlenk, E.~Villa   \\
{\em Program Title: } py\textsc{SecDec}                                    \\
{\em Developer's repository:} \url{https://github.com/gudrunhe/secdec} \\
{\em Online documentation:} \url{https://secdec.readthedocs.io} \\
%{\em Journal Reference:}                           \\
%{\em Program Files doi:}                              \\
 {\em Licensing provisions: GNU Public License v3}                                   \\
  %enter "none" if CPC non-profit use license is sufficient.
{\em Programming language:} Python, \textsc{Form}, C++, \textsc{Cuda}     \\
{\em Computer:} from a single PC/Laptop to a cluster, depending on the
problem; if the optional GPU support is used, \textsc{Cuda} compatible hardware is required.\\
  %Computer(s) for which program has been designed.
{\em Operating system: } Unix, Linux                                      \\
  %Operating system(s) for which program has been designed.
{\em RAM:} hundreds of megabytes or more, depending on the complexity of the problem                                              \\
  %RAM in bytes required to execute program with typical data.
  %{\em Number of processors used:}                              \\
  %If more than one processor.
  %{\em Supplementary material:}                                 \\
  % Fill in if necessary, otherwise leave out.
 \\
%{\em PACS:}      
%12.38.Bx, 
%02.60.Jh, 	
%02.70.Wz \\
{\em Classification:}                                         
  4.4 Feynman diagrams, 
  5 Computer Algebra, 
  11.1 General, High Energy Physics and Computing.\\
{\em External routines/libraries:}
\textsc{GSL}~[1],
\textsc{NumPy}~[2],
\textsc{SymPy}~[3],
\textsc{Nauty}~[4],
\textsc{Cuba}~[5],
\textsc{Form}~[6],
\textsc{GiNaC}~[7]; optionally
\textsc{Normaliz}~[8]. \\
%{\em Subprograms used:}                                       \\
%Fill in if necessary, otherwise leave out.
%{\em Catalogue identifier of previous version:}*              \\
%Only required for a New Version summary, otherwise leave out.
{\em Journal reference of previous version:} \href{https://doi.org/10.1016/j.cpc.2019.02.015}{Comput. Phys. Commun. 240 (2019) 120--137}.\\
{\em Does the new version supersede the previous version?:} yes  \\
{\em Nature of the problem:}\\
 Expansion of Feynman integrals in different kinematic limits,
 regularisation of ultraviolet, infrared and spurious singularities, 
  numerical integration in the presence of integrable singularities 
  (e.g. kinematic thresholds). \\
  {\em Solution method:}\\
  After specification of the desired kinematic limit by the user
  (definition of a smallness parameter), 
  the regions contributing to the integral are determined and
  expansion in the smallness parameter up to the desired order is performed.
 Extraction of singularities in the dimensional regularization
 parameter as well as in analytic regulators for potential spurious
 singularities is done using sector decomposition. 
 This leads to a Laurent series in the regularization
 parameters, 
 where the coefficients are finite integrals over the unit hypercube.
 Integrable singularities are handled by 
 choosing a suitable integration contour in the complex plane, in an
 automated way.
The integrals expanded in the different regions are summed and evaluated numerically.
    \\
%{\em Reasons for the new version:}*\\
%{\em Summary of revisions:}*\\
{\em Restrictions:} Depending on the complexity of the problem, limited by 
memory and CPU/GPU time.\\
%   \\
{\em Running time:}
Between a few seconds and several days, depending on the comp\-lexity of the problem.\\
  %Give an indication of the typical running time here.
{\em References:}
\begin{refnummer}
\item M. Galassi et al,
    GNU Scientific Library Reference Manual.
    \texttt{ISBN:0954612078},
    \url{http://www.gnu.org/software/gsl/}.
\item C.~R.~Harris, K.~J.~Millman, S.~J.~van~der~Walt, et al,
    Array programming with NumPy,
    Nature \textbf{585} (2020) 357--362.
    \texttt{\href{https://doi.org/10.1038/s41586-020-2649-2}{doi:10.1038/s41586-020-2649-2}},
    \url{http://www.numpy.org/}.
\item A.~Meurer, et al.,
    SymPy: symbolic computing in Python,
    PeerJ\ Comp.\ Sci. \textbf{3} (2017) e103.
    \texttt{\href{https://doi.org/10.7717/peerj-cs.103}{doi:10.7717/peerj-cs.103}},
    \url{http://www.sympy.org/}.
\item B.~D.~McKay and A.~Piperno,
    Practical graph isomorphism, II,
    J.\ Symb.\ Comput.\ \textbf{60} (2014) 94--112.
    \texttt{\href{https://doi.org/10.1016/j.jsc.2013.09.003}{doi:10.1016/j.jsc.2013.09.003}},\\
    \url{http://pallini.di.uniroma1.it}.
\item T.~Hahn,
    CUBA: A Library for multidimensional numerical integration,
    Comput.\ Phys.\ Commun.\ \textbf{168} (2005) 78.
    \texttt{\href{https://arxiv.org/abs/hep-ph/0404043}{arXiv:hep-ph/0404043}},\\
    \url{http://www.feynarts.de/cuba/}.
\item J.~Kuipers, T.~Ueda and J.~A.~M.~Vermaseren,
    Code Optimization in FORM,
    Comput.\ Phys.\ Commun.\ \textbf{189} (2015) 1.
    \texttt{\href{https://arxiv.org/abs/1310.7007}{arXiv:1310.7007}},\\
    \url{http://www.nikhef.nl/~form/}.
\item C.~W.~Bauer, A.~Frink, and R.~B.~Kreckel,
    Introduction to the GiNaC framework for symbolic computation within the C++ programming language,
    J.\ Symb.\ Comput.\ \textbf{33} (2002) 1--12.
    \texttt{\href{https://arxiv.org/abs/cs/0004015}{arXiv:cs/0004015}},
    \url{https://www.ginac.de/}.
\item W.~Bruns, B.~Ichim, B. and T.~R{\"o}mer, C.~S{\"o}ger,
    Normaliz. Algorithms for rational cones and affine monoids.
    \url{http://www.math.uos.de/normaliz/}.
\end{refnummer}
\end{small}

%\hspace{1pc}
%{\bf LONG WRITE-UP}

\section{Introduction}
\label{sec:intro}
Precision calculations play a leading role in high energy physics, at
the Large Hadron Collider and in view of other current and future experiments.
The complexity of the calculation of higher order corrections in
perturbation theory however often prohibits a direct analytic evaluation of
the corresponding expressions.
However, approximations of the integrals in certain limits
can often be easier to calculate and nonetheless sufficiently accurate
for phenomenological purposes.
Prominent examples are inclusive Higgs production in gluon fusion at N$^3$LO in
the large-$m_t$ expansion, with~\cite{Anastasiou:2015ema} and without~\cite{Mistlberger:2018etf} the threshold approximation,
$b$-quark effects in $H+$jet production~\cite{Lindert:2017pky,Caola:2018zye},
the high-energy expansion in Higgs-boson pair
production~\cite{Mishima:2018olh,Davies:2018qvx,Davies:2019dfy},
$gg\to ZZ$~\cite{Davies:2020lpf} and $gg\to ZH$~\cite{Davies:2020drs} production, 
combined threshold plus large-$m_t$ expansions in $gg\to HH$~\cite{Grober:2017uho},
$gg\to ZZ$~\cite{Grober:2019kuf} and  $gg\to ZH$~\cite{Alasfar:2021ppe}, or expansions in the limit of small external masses~\cite{Wang:2021rxu}.

A systematic approach to the expansion of Feynman integrals in various
limits, characterised by a scaling parameter which is small in each limit, is
the so-called {\em expansion by regions}.
The method of expansion by regions has been pioneered in
Refs.~\cite{Smirnov:1991jn,Beneke:1997zp,Smirnov:1998vk}, where it was formulated
in terms of the momenta involved in a loop integral.
Later it also has been formulated in Feynman parameter space, where it
allows a geometric interpretation~\cite{Smirnov:1999bza,Pak:2010pt,Ananthanarayan:2018tog,Ananthanarayan:2020ptw}.
A remarkable feature of this method is that in most cases, it allows to reproduce the 
result for the original integral if  all regions are
identified correctly and summed up, after  having performed the integrations
 in the various regions  over the full integration domain.
In Ref.~\cite{Jantzen:2011nz}, the relations between the expanded
integrals and the original integral have been investigated in detail
with the aim to clearly identify the conditions under which the original integral
can be fully reproduced by summing up the individual regions. It was shown that
overlap contributions usually yield scaleless integrals which can be
set to zero if they are regulated appropriately.

Still, there are open issues with the strategy of regions, in particular where terms of differing sign appear in the graph polynomials, violating the assumption that a Euclidean region exists.
This can happen for example in Glauber or potential regions, where in simple cases a split of the integration region combined with a change variables may solve the issue~\cite{Jantzen:2012mw}, but a general solution is difficult to find.
Ref.~\cite{Semenova:2018cwy} contains the following statement about the strategy of regions:
``Although this strategy certainly looks suspicious for mathematicians, it was successfully applied in numerous calculations.
It has the status of experimental mathematics and should be applied
with care''.

The method also has been implemented in the code \textsc{Asy2.m}~\cite{Jantzen:2012mw}, which is part of the program \textsc{Fiesta}~\cite{Smirnov:2013eza,Smirnov:2015mct}.
Recently, a method using Groebner basis techniques to solve the Landau equations has been suggested to identify regions for multi-scale integrals in an algorithmic way, implemented
in the Mathematica code \textsc{Aspire}~\cite{Ananthanarayan:2018tog,Ananthanarayan:2020ptw}.
An alternative Feynman parameter space based expansion method using Mellin-Barnes techniques~\cite{Pilipp:2008ef} is implemented in the recently published program \textsc{HepLib}~\cite{Feng:2021kha}.
Another approach to asymptotic expansions based on loop-tree-duality is presented in Ref.~\cite{Plenter:2020lop}.

In this paper, we present an implementation of the method of  expansion
by regions in \pysecdec~\cite{Borowka:2017idc,Borowka:2018goh}.
Such automated expansions can be particularly useful in cases where the aim is not to
reproduce the full original integral, but to get a result in certain kinematic limits.
Our implementation is based on the geometric formulation of expansion by regions.

In addition, we present major new features, including the possibility to calculate amplitudes (weighted sums of integrals).
We have also implemented an improved algorithm to adjust the parameters defining the deformation of the integration contour,
such that the deformation parameters $\lambda_i$ are automatically decreased if the default values would lead to an invalid contour
for some phase space points, thereby removing one of the most frequent issues encountered in previous versions of the code.

Furthermore, we give some analytic insights about the criteria for the absence of overlap terms in multiple expansions,
which, to the best of our knowledge, were not present in the literature before within the geometric formulation of expansion by regions.

This article is structured as follows. In Section~\ref{sec:regions} we describe parametric representations of multi-loop Feynman integrals and briefly introduce the concept of expansion by regions.
Section~\ref{sec:ebralg} is dedicated to the geometric formulation of expansion by regions.
In Section~\ref{sec:newfeatures} we list the new features of the release, while Section~\ref{sec:usage} describes the usage of
expansion by regions in \pysecdec{} and gives additional information about timings and convergence behaviour in the context of asymptotic expansions.
Section~\ref{sec:examples} contains several examples demonstrating the new features.

\section{Feynman integrals and an example for expansion by regions}
\label{sec:regions}

\subsection{Feynman integrals}

We consider Feynman integrals with $L$ loops in $D$ space-time dimensions,
\begin{align}\label{eq:Feynman}
I(\vec{\nu})  &=
\int_{-\infty}^\infty \left(\prod\limits_{l=1}^{L} \frac{d^Dk_l}{i\pi^{\frac{D}{2}}}\right)
\prod\limits_{j=1}^{N} 
\frac{1}{P_{j}^{\nu_j}(\{k\},\{p\},m_j^2)}\;, 
\end{align}
where the propagators $P_{j}$ are of the form $P_{j}^{\nu_j}(\{k\},\{p\},m_j^2)=q_j^2-m_j^2+i\delta$, with $q_j$ being a linear combination of loop momenta $k$ and external momenta $p$.
Introducing Feynman parameters based on the relation
\begin{align}\label{eq:Feynmanpara}
  \prod\limits_{j=1}^{N} \frac{1}{P_{j}^{\nu_j}}&
  =\frac{\Gamma(N_\nu)}{\prod\limits_{j=1}^N\Gamma(\nu_j)}\int_0^\infty \left(\prod\limits_{j=1}^N\,dx_j\,x_j^{\nu_j-1}\right)
                                            \frac{\delta(1-\sum_{i=1}^N x_i)}{\left( \sum_{j=1}^Nx_jP_j \right)^{N_\nu}} \;,\; N_\nu=\sum_{j=1}^N\nu_j   \;,                                              
\end{align}
the integral has the form
\begin{align}
I(\vec{\nu}) &= \frac{\Gamma(N_\nu)}{\prod\limits_{j=1}^{N}\Gamma(\nu_j)}\int_0^\infty \left(\prod\limits_{j=1}^{N}\,dx_j\,x_j^{\nu_j-1}\right)
 \, \delta(1-\sum_{i=1}^N x_i)\nn\\
&\times \int_{-\infty}^\infty d\kappa_1\dots d\kappa_L 
\left[ 
       \sum\limits_{j,l=1}^{L} k_j\cdot k_l \, M_{jl} - 
      2\sum\limits_{j=1}^{L} k_j\cdot Q_j +J +i\delta\right]^{-N_\nu}\;,\nn
\end{align}
where $\kappa_l=d^Dk_l/(i\pi^{\frac{D}{2}})$.
Momentum integration leads to a form depending on the Symanzik polynomials $\mathcal{U}$ and $\mathcal{F}$,
\begin{align}
  I(\vec{\nu})&= \frac{(-1)^{N_{\nu}}}{\prod_{j=1}^{N}\Gamma(\nu_j)}\Gamma(N_{\nu}-LD/2)
 \int\limits_{0}^{\infty} \left(\prod\limits_{j=1}^{N}dx_j\,\,x_j^{\nu_j-1}\right)\,\delta(1-\sum_{l=1}^N x_l)\,
                \frac{\mathcal{U}^{N_{\nu}-(L+1) D/2}}{{\mathcal F}^{N_\nu-L D/2}}\,,\nn\\
                &\nn\\
{\cal U}&=\det(M)\; , \;
{\cal F}=\det(M)\,\left[ \sum_{i,j=1}^L Q_i \left(M^{-1}\right)_{ij}Q_j-J-i\,\delta  \right]\;.
\label{eq:feynint}
\end{align}
The second Symanzik polynomial, ${\cal F}$, contains the kinematic invariants the result for the integral will depend on.

A closely related representation has been derived by Lee and Pomeransky~\cite{Lee:2013hzt}:
\begin{equation}
 I(\vec{\nu})  =\frac{(-1)^{N_\nu}\Gamma\left(D/2\right)}{\Gamma\left(\left(L+1\right)D/2-N_\nu\right)\prod_{j}\Gamma\left(\nu_{j}\right)}\int\limits _{0}^{\infty}\left(\prod_{j=1}^Ndz_{j}\,z_{j}^{\nu_{j}-1}\right)({\cal U}+{\cal F})^{-D/2}\,.\label{eq:LeePom}
\end{equation}
It can be proven to be equivalent to the form in Eq.~(\ref{eq:feynint})
by inserting $1=\int_0^\infty d\eta\,\delta(\eta-\sum_{j=1}^{N}z_j)$, substituting
$z_j=\eta\,x_j$ for $ j=1,\ldots, N$ and identifying the corresponding $\Gamma$-functions.

In cases where the kinematic invariants are not of the same order of magnitude, it can be beneficial to expand the integrand in a small parameter $t$, namely the ratio of appropriate kinematic invariants, for instance $t=s_{ij}/m^2$ in a large mass expansion.
The individual terms of such an expansion then may be easier to evaluate.
However, if we want to calculate an integral with corrections up to some order in a small parameter $t$, it is not always possible to just Taylor expand the integrand.
The integrand might not have a convergent Taylor series in the whole integration domain and we may need different series expansions in different regions of the integration domain.
Therefore a more careful strategy is required to perform valid expansions.
An essential feature of Feynman integrals used for this purpose is their homogeneity property under rescaling of the integration parameters, which we will discuss in the following section.

\subsection{Cheng-Wu theorem}
The Cheng-Wu theorem~\cite{Cheng:1987ga} states that integrals of the type
\begin{align}
  \int_0^\infty \left(\prod\limits_{j=1}^N\,dx_j\right) f(\{x_i\})
    \,\, \delta(1-\sum_{i=1}^N x_i)\;,
\end{align}
are invariant under rescalings of the parameters in the  delta-function of the form
$\delta(1-\sum_{i=1}^N x_i) \rightarrow \delta(1-\sum_{i=1}^N a_i x_i)$, with $a_i \geq 0$ and at least one $a_i \neq 0$,
provided that the function $f$ together with the integration measure is invariant under rescaling of all integration variables by a constant $\eta$, i.e. $(\prod\,dx_j) f(\{x_i\}) = (\prod\,dx_j^\prime) f(\{x_i^\prime\})$ with $x_i^\prime= \eta x_i$ for all $i$.
Feynman integrals do have this property.

One way to prove the Cheng-Wu-theorem is to perform the primary sector decomposition as done in
Refs.~\cite{Binoth:2000ps,Heinrich:2008si}, and observe that the primary sectors will be the same,
regardless of the coefficients in the delta-function argument. The explicit proof can be found in Appendix~\ref{sec:appendixWu}.

The Cheng-Wu theorem allows us to perform the changes of variables used in expansion by regions without having to consider changes of the coefficients in the argument of the delta-function.

\subsection{An example in momentum space}
\label{sec:momspace}

The various regions that can occur depend on the considered integral. In general there is a hard region where all components of the loop momentum are much bigger than the parameter we are expanding in.
In addition, there can be various collinear regions, a soft region, an ultrasoft region, and other regions.
For a discussion of these regions in momentum space we refer e.g. to Ref.~\cite{Becher:2014oda}.
In more involved cases, also
Glauber regions or potential regions may need to be considered, related for example to forward scattering or threshold expansions.
The latter lead to polynomials composed of monomials with both positive and negative coefficients, which
are briefly discussed in Section~\ref{sec:Glauber}.

\begin{figure}[htb]
  \centering
  {\begin{tikzpicture}
	\begin{pgfonlayer}{nodelayer}
		\node [style=none] (0) at (-1.5, 0) {};
		\node [style=dot] (1) at (-1, 0) {};
		\node [style=dot] (2) at (1, 0) {};
		\node [style=none] (3) at (1.5, 0) {};
		\node [style=none] (4) at (1, 0.75) {$m$};
		\node [style=none] (5) at (-1.25, 0.25) {$s$};
	\end{pgfonlayer}
	\begin{pgfonlayer}{edgelayer}
		\draw [style=edge, style=dot1, bend left=75, looseness=1.25] (1) to (2);
		\draw [style=external edge] (0.center) to (1);
		\draw [style=external edge] (2) to (3.center);
		\draw [style=massless edge, bend right=75, looseness=1.25] (1) to (2);
	\end{pgfonlayer}
\end{tikzpicture}}
  \caption{%
    One loop bubble with a squared massive propagator.
    Dashed lines denote massless propagators.
    The ``dot'' means that the corresponding propagator occurs with power two in the integral.}
  \label{fig:dotted_bubble}
\end{figure}
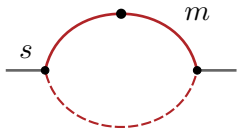

To explain the basic principle, let us consider a simple example in momentum space, discussed in more detail in Ref.~\cite{Jantzen:2011nz}.
We consider the large-momentum expansion of the integral
\be
I_2 = \int \mathcal{I}_2=\mu^{2\eps}\int d\kappa\frac{1}{(k+p)^2(k^2-m^2)^2}\;,
\ee
depicted in Fig.~\ref{fig:dotted_bubble}, see also the example in Section~\ref{sec:usage}.

We will expand the integrand in $m^2/p^2$ in the limit $|p^2|\gg m^2$.
It is helpful to consider the first component of $k$ to be imaginary and to perform a Wick rotation to momenta in Euclidean space. In this way the smallness of a squared momentum means smallness of its components.
We can introduce a smallness parameter $z$, in terms of which a series expansion is performed.
For the hard region, where $m^2$ is small compared to $|k^2|$, we introduce $m^2\rightarrow z\,m^2$, expand about $z=0$ and then set $z=1$.
\be
(h): \quad \frac{1}{(k+p)^2(k^2-m^2)^2}\to \frac{1}{(k+p)^2\,(k^2)^2}\left( 1+2\frac{m^2}{k^2}+\ldots\right).
\label{eq:hard}
\ee
This expansion is valid whenever $|k^2| > |m^2|$. However, since the
integration domain also contains points that do not satisfy this
condition, we need another region. This is the soft region, where we
introduce $z$ by $k^2\rightarrow z\,k^2$, $p\cdot k\rightarrow z^{1/2}\,p\cdot k$, $m^2\rightarrow z\,m^2$ and expand about $z=0$, after which we set $z=1$.
\be
  (s): \quad \frac{1}{(k+p)^2(k^2-m^2)^2}\to \frac{1}{p^2\,(k^2-m^2)^2}\left( 1-\frac{k^2+2p\cdot k}{p^2}+\ldots\right).
  \label{eq:soft}
  \ee
The resulting expansion is valid whenever $k^2+2k\cdot p < p^2$. This means that this expansion covers the region where the hard region expansion, \eqn{eq:hard}, does not converge.

By introducing a boundary $\Lambda$ with $m^2 \ll \Lambda^2 \ll | p^2 |$ one can split the integration region into two parts $D_h = \left\{ k \in \mathbb{R}^D \mid |k^2| \geq \Lambda^2 \right\}$ and $D_s = \left\{ k \in \mathbb{R}^D \mid |k^2| < \Lambda^2 \right\}$ in which the respective expansions converge:
\be
I_2 = \int_{D_h}\mathcal{I}_2^{(h)} +\int_{D_s}\mathcal{I}_2^{(s)} =\left[ \int\mathcal{I}_2^{(h)} - \int_{D_s}\mathcal{I}_2^{(sh)}\right] +\left[\int\mathcal{I}_2^{(s)} - \int_{D_h}\mathcal{I}_2^{(hs)}\right].
\label{eq:loopexebr1}
\ee
Here the integration regions of both expansions are again extended over the whole integration domain. To compensate for this, additional terms are subtracted.
The extra contributions are also expanded in the expansion parameter which converges in their respective integration domains.
In cases where the expansions commute, i.e. $\mathcal{I}_2^{(hs)} = \mathcal{I}_2^{(sh)}$, one can combine the extra contributions, resulting in
\be
I_2 = \int\mathcal{I}_2^{(h)} + \int\mathcal{I}_2^{(s)} - \int\mathcal{I}_2^{(hs)}.
\label{eq:loopexebr2}
\ee
The term involving the double expansion turns out to be a scaleless integral, which vanishes if properly regularised, in our example by dimensional regularisation.
To see this, if we take the expansion in the hard region, expression \eqn{eq:hard},  and expand it again in the soft region, we get
\be
  (hs): \frac{1}{(k+p)^2(k^2-m^2)^2}\to \frac{1}{p^2\,(k^2)^2} \left( 1-\frac{k^2+2p\cdot k}{p^2}+\ldots\right) \left( 1+2\frac{m^2}{k^2}+\ldots\right).
\label{eq:overlap}
\ee
Expanding the expression of the soft region, \eqn{eq:soft}, again in
the hard region leads to the same result, which shows that the hard and soft expansion commute.
These are scaleless integrals to all orders in the expansion, and therefore we can ignore them.

Integrating the leading contributions in the soft and hard regions gives
\begin{align}
I_2^{(h)}&=\frac{1}{p^2}\left[ -\frac{1}{\eps}+\ln\left( \frac{-p^2}{\mu^2}\right) \right]+{\cal O}(\eps, m^2/p^2)\;,\label{eq:epsIR}\\ 
I_2^{(s)}&=\frac{1}{p^2}\left[ \frac{1}{\eps}-\ln\left( \frac{m^2}{\mu^2}\right) \right]+{\cal O}(\eps,m^2/p^2)\;,\label{eq:epsUV}
\end{align}
such that
\be
I_2=I_2^{(h)}+I_2^{(s)}=\frac{1}{p^2}\,\ln\left( \frac{-p^2}{m^2}\right)+{\cal O}(\eps,m^2/p^2)\;.
\ee
Note that in both the soft and the hard region we encounter spurious poles that do not exist in the original integral.
The pole in \eqn{eq:epsIR} is of infrared nature, while the pole in \eqn{eq:epsUV} is of UV nature, and
\eqn{eq:overlap} can be written as
\be
I_2^{(hs)}=\frac{1}{p^2}\left( \frac{1}{\eps_{UV}}- \frac{1}{\eps_{IR}}\right)\;.
\ee
Therefore, formally, the combinations $I_2^{(h)}-I_2^{(hs)}$ and $I_2^{(s)}-I_2^{(hs)}$ are separately IR and UV finite.
In more general cases, the spurious poles are not always regulated by dimensional regularisation.
Extra regulators must be introduced, which is most conveniently done by using so-called ``analytic regulators''~\cite{Smirnov:1997gx,Becher:2010tm,Becher:2014oda}, where the propagators get extra powers $\alpha$, and the $\alpha$-dependence should cancel in the final result. An example for such a case is given by the high-energy expansion of a one-loop box diagram with a massive loop, discussed in Section~\ref{sec:examples} in the context of \pysecdec{} and analysed in great detail in Ref.~\cite{Mishima:2018olh}.

For many complicated examples it has been shown in the
literature~\cite{Smirnov:1991jn,Beneke:1997zp,Manohar:2006nz,Jantzen:2011nz,Mishima:2018olh}
that one can integrate each expansion over the whole integration domain to get the correct result.

In the following, the method of expansion by regions in Feynman parameter space is reviewed. The geometric approach to determine all the relevant regions is discussed and insights about the validity of the method are presented.

\section{Geometric formulation of expansion by regions}
\label{sec:ebralg}

The geometric method has been introduced in~\cite{Pak:2010pt} and is also described in~\cite{Smirnov:2012gma,Ananthanarayan:2018tog}.
It uses many of the ideas also used in the geometric formulation of sector decomposition~\cite{Bogner:2007cr,Kaneko:2009qx,Kaneko:2010kj}, implemented in {\sc SecDec-3}~\cite{Borowka:2015mxa,Schlenk:2016epj,Schlenk:2016cwf} and \pysecdec~\cite{Borowka:2017idc,Borowka:2018goh}.

\subsection{Geometric method of determining the regions}
\label{sec:geo}

We discuss the expansion of parametric integrals over polynomials of the form 
\begin{equation}
P(\mathbf{x},t) = \sum_{i=1}^{m} c_i x_1^{p_{i,1}} \!{}_{\dots}\ x_N^{p_{i,N}} t^{p_{i,N+1}} 
\label{eq:poly}
\end{equation}
around small values of
the so-called {\em smallness parameter} $t$, such as the quantity $m^2/p^2$ in the large-momentum expansion in the example above. 
Feynman integrals can be brought into this form by using the Lee-Pomeransky representation, \eqn{eq:LeePom},
leading to
\begin{equation}
I = \int_0^{\infty}\,\frac{\mathrm{d}\mathbf{x}}{\mathbf{x}} \mathbf{x}^{\bm{\nu}} t^{\nu_{N+1}} \left[ \sum_{i=1}^m c_i \mathbf{x}^{\mathbf{p}_i} t^{p_{i,N+1}} \right]^b \;,
\label{eq:int}
\end{equation}
where $b=-D/2$ and $\mathbf{x}^{\mathbf{a}} =\prod_{j=1}^N x_j^{a_j}$.
The exponents can be organised into $(N+1)$-dimensional vectors $\mathbf{p}'_i \equiv \left(\mathbf{p}_i, p_{i,N+1} \right)$, $\bm{\nu}' \equiv \left(\bm{\nu}, \nu_{N+1} \right)$.
We assume that all exponent vectors $\mathbf{p}'_i$ are restricted to the lattice $\mathbb{Z}^{N+1}$.
For all polynomial coefficients we require $c_i \geq 0$.
This constraint can be relaxed, as will be discussed in Section~\ref{sec:Glauber}.

The goal of the expansion by regions algorithm is to obtain an expansion of $I$ in terms of the smallness parameter $t$, where the coefficients of the expansion are integrals depending only on the Feynman parameters and not on
$t$. For each region we can perform a change of variables of the form
\begin{align}
t&\rightarrow z t\;,\;
x_j\rightarrow z^{v_{j}} x_j\;,
\label{eq:z-method}
\end{align}
expand about $z=0$ and then set $z=1$.
The vector $\mathbf{v}=(v_1,\dots,v_N,1)$ describes the rescaling applied to $(\mathbf{x},t)$ and is called the {\em region vector}.
Up to a change of variables, this method is equivalent to the replacement
\begin{align}
%t&\rightarrow  t\;,\;
x_j\rightarrow t^{v_{j}} x_j\;,
\label{eq:t-method}
\end{align}
leaving $t$ unchanged, and series expanding in $t$, as proven in Appendix~\ref{sec:appendix1}.
The equivalence of methods (\ref{eq:z-method}) and (\ref{eq:t-method}) allows us to get easily a power series in $t$ by expanding in $z$. We will call method (\ref{eq:z-method}) the {\em z-method} and method (\ref{eq:t-method}) the {\em t-method} in the following.

In \eqn{eq:int}  we consider the Lee-Pomeransky representation for simplicity, but the same methods can be applied to integrals in the Feynman representation as given in Eq.~(\ref{eq:feynint}), containing a product of two polynomials $U$ and $F$
and a delta-function. The resulting region vectors of the two different parametrisations are the same up to constant shifts with multiples of $(\vec{1},0)$~\cite{Semenova:2018cwy,AndresThesis}.

The notion of the {\em Newton polytope}  of a polynomial is of crucial importance in the geometric formulation of expansion by regions, since it allows us to find all the relevant regions.
The Newton polytope can be determined as the convex hull of the exponent vectors:
\begin{equation}
\Delta' = \mathrm{convHull}\left(\mathbf{p}'_1,\mathbf{p}'_2,\dots\right)=\left\{\sum_j\alpha_j \mathbf{p}'_j\;\Big|\,  \alpha_j\geq 0 \wedge \sum_j\alpha_j =1\right\}\;.
\end{equation}
Here $\Delta'$ denotes the $(N+1)$-dimensional Newton polytope which includes the smallness parameter $t$, while $\Delta$ is the $N$-dimensional Newton polytope involving only the integration parameters.

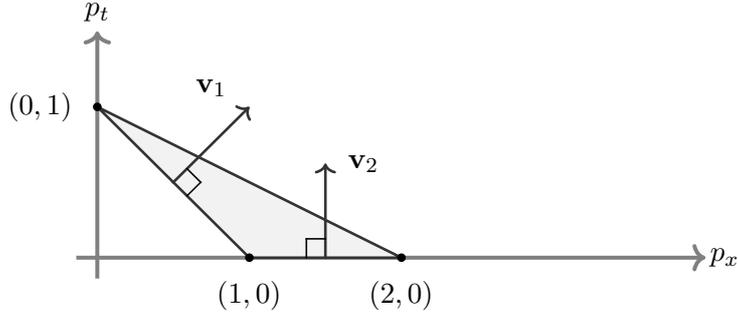
\begin{figure}[h]
    \centering
    {\begin{tikzpicture}
	\begin{pgfonlayer}{nodelayer}
		\node [style=none] (0) at (0, -0.25) {};
		\node [style=none] (1) at (8, 0) {};
		\node [style=none] (2) at (0, 3) {};
		\node [style=dot] (3) at (0, 2) {};
		\node [style=dot] (4) at (2, 0) {};
		\node [style=dot] (5) at (4, 0) {};
		\node [style=none] (6) at (3, 0) {};
		\node [style=none] (7) at (1, 1) {};
		\node [style=none] (8) at (2, 2) {};
		\node [style=none] (9) at (3, 1.25) {};
		\node [style=none] (10) at (-0.75, 2) {$(0,1)$};
		\node [style=none] (11) at (2, -0.5) {$(1,0)$};
		\node [style=none] (12) at (4, -0.5) {$(2,0)$};
		\node [style=none] (13) at (0, 3.25) {$p_t$};
		\node [style=none] (14) at (8.25, 0) {$p_x$};
		\node [style=none] (15) at (3.5, 1.25) {$\mathbf{v}_2$};
		\node [style=none] (16) at (1.5, 2.25) {$\mathbf{v}_1$};
		\node [style=none] (17) at (1.36, 1) {};
		\node [style=none] (18) at (1.18, 0.82) {};
		\node [style=none] (19) at (1.18, 1.18) {};
		\node [style=none] (20) at (2.75, 0) {};
		\node [style=none] (21) at (2.75, 0.25) {};
		\node [style=none] (22) at (3, 0.25) {};
		\node [style=none] (23) at (-0.25, 0) {};
	\end{pgfonlayer}
	\begin{pgfonlayer}{edgelayer}
		\draw [style=thick line, ->, gray] (0.center) to (2.center);
		\draw [style=thick line, ->, gray] (23.center) to (1.center);
		\draw [style=line, line join=round, fill={black!5!white}] (3.center)
			 to (4.center)
			 to (5.center)
			 to cycle;
		\draw [style=line, ->] (6.center) to (9.center);
		\draw [style=line, ->] (7.center) to (8.center);
		\draw [style=thin line] (19.center) to (17.center);
		\draw [style=thin line] (17.center) to (18.center);
		\draw [style=thin line] (20.center) to (21.center);
		\draw [style=thin line] (21.center) to (22.center);
	\end{pgfonlayer}
\end{tikzpicture}}
    \caption{Newton polytope for $P(x,t) = t+x+x^2$, together with the directions of the region vectors (drawn as normal vectors to the facets in positive $t$-direction).}
    \label{fig:polytope_simple}
\end{figure}

Alternatively, a convex polytope can also be described as the intersection of half spaces
\begin{equation}
\Delta' = \bigcap_{f\in F} \left\{ \mathbf{m}\in\mathbb{R}^{N+1} \mid \langle \mathbf{m},\mathbf{n}_f\rangle + a_f \geq 0 \right\}\;,
\end{equation}
where $F$ is the set of polytope facets with inward-pointing normal vectors $\mathbf{n}_f$, $\langle \mathbf{m},\mathbf{n}_f\rangle$ is the scalar product of $\mathbf{m}$ and $\mathbf{n}_f$, and $a_f\in \mathbb{Z}$.
The subset of facets with normal vectors pointing in the positive $t$-exponent direction is $F^+ = \left\{ f \in F \mid (\mathbf{n}_f)_{N+1} > 0\right\}$. These can be used as input for the change of variables introduced above:
\begin{align}
t &\rightarrow \prod_{f \in F^+} z_f^{(\mathbf{n}_f)_{N+1}} t\;,\nn\\
x_i &\rightarrow  \prod_{f \in F^+} z_f^{(\mathbf{n}_f)_i} x_i\;,
\label{eq:trafo}
\end{align}
where for each region $f$, $\mathbf{n}_f$ is the region vector $\mathbf{v}$ introduced in~\eqref{eq:z-method}.
An example of a Newton polytope for the polynomial 
\be
P(x,t) = t+x+x^2
\label{eq:examplepoly}
\ee
is shown in Fig.~\ref{fig:polytope_simple}.
There are two regions, defined by the region vectors $\mathbf{v}_1=(1,1)$ and $\mathbf{v}_2=(0,1)$.

The transformations \eqref{eq:trafo} lead us to the following form of~\eqref{eq:int}:
\begin{equation}
I = \left(\prod_{f\in F^+} z_f^{\langle \mathbf{n}_f,\bm{\nu}'\rangle - b a_f}\right) \int_0^{\infty}\,\frac{\mathrm{d}\mathbf{x}}{\mathbf{x}} \mathbf{x}^{\bm{\nu}} t^{\nu_{N+1}}  \left[ \sum_i c_i \mathbf{x}^{\mathbf{p}_i} t^{p_{i,N+1}} \prod_{f\in F^+} z_f^{\langle \mathbf{n}_f,\mathbf{p}'_i\rangle + a_f}  \right]^b.
\label{eq:intrescaled}
\end{equation}

According to the geometric expansion by regions algorithm, the original integral can be approximated by
expanding in $z_f$ in each region while setting the other $z_{f'}$ to one,
and then setting  $z_f$ to one after the expansion in region $f$.
This leads to
\begin{equation}
I = \sum_{f\in F^+} I_f\;,
\label{eq:ebrgen}
\end{equation}
where $I_f$ is the expansion of \eqref{eq:intrescaled} in $z_f$ with all $z$ set to one after the expansions.
The $I_{f}$ are integrated over the whole integration domain.

Here we also want to point out that the set of facets $F^+$ of $\Delta'$ induces a subdivision $\mathrm{SD}(\Delta)$ of the $N$-dimensional Newton polytope $\Delta$.
The subdivision is obtained by projecting the facets in $F^+$ from $\mathbb{R}^{N+1}$ to the $(\mathbb{R}^N,0)$ plane along the $\hat{\mathbf{e}}_{N+1}$ direction.
In the example of Fig.~\ref{fig:polytope_simple}  the set $F^+$ contains two facets $f_1$ and $f_2$ with normal vectors $\mathbf{v}_1$ and $\mathbf{v}_2$.
The Newton polytope $\Delta$ of the unexpanded integral is the interval $[0,2] \in \mathbb{R}$.
Projecting the $F^+$ facets subdivides $\Delta = [0,2]$ into two intervals $[0,1]$ and $[1,2]$ corresponding to the facets $f_1$ and $f_2$ respectively.
Two more examples of subdivisions are shown in Fig.~\ref{fig:subdivex}.

The basic foundation for the validity of this approach was given by
Jantzen in Ref.~\cite{Jantzen:2011nz}. There he proved algebraically a
relation between the original integral and its expansions. 
The relation is a generalisation of \eqn{eq:loopexebr2}, which is only valid for expansions with two commuting regions, to an arbitrary number of regions.
The general relation can be written schematically as
\begin{equation}
I = \left\{\textrm{single exp.}\right\} - \left\{\textrm{double exp.}\right\} + \dots - (-1)^{N} \left\{N\textrm{-fold exp.}\right\} + I_{nc}\;,
\label{eq:Jantzen}
\end{equation}
where $N$ is the number of regions, the curly brackets represent sums of commuting expansions integrated over the whole integration domain, and $I_{nc}$ is a complicated sum of multiple non-commuting and commuting expansions integrated only in the convergence domain of the non-commuting expansions. 
Jantzen's formula clearly reduces to the simple expansion by regions relation equivalent to \eqn{eq:ebrgen} when the terms with multiple expansions and $I_{nc}$ vanish.

In Sections~\ref{sec:convdomain} and \ref{sec:facets_full_domain} the convergence domain of different expansions is discussed and the choice of facet normal vectors as region vectors is justified.
In Section~\ref{sec:multi_exp_vanish} we will show within the geometric approach to expansion by regions why integrals with multiple expansions vanish if dimensional or analytic regularisation is used.
Finally in Section~\ref{sec:commuting} we show that for a large class of limits of Feynman integrals the non-commuting term $I_{nc}$ in \eqn{eq:Jantzen} vanishes.

\subsubsection{Convergence domain of an expansion with a region vector}
\label{sec:convdomain}

We first discuss the domain of convergence for an expansion done with a general region vector $\mathbf{v}$.
We consider a polynomial of the form of \eqn{eq:poly}.
As already mentioned, each term of the polynomial can be represented by a vector $\mathbf{p}'_i$ in $(N+1)$-dimensional space. All the terms together constitute the Newton polytope of the polynomial.

Furthermore, we define an $N$-dimensional vector $\mathbf{u}$ by $x_i=t^{u_i}$ for each point $\mathbf{x}$ in the integration domain, and an associated $(N+1)$-dimensional vector $\mathbf{u}'=\left(\mathbf{u},1\right)$.
The change from the integration variables $\mathbf{x}$ to $\mathbf{u}$ or $\mathbf{u}'$ can be viewed as a change of variables mapping the original integration domain $\mathbb{R}^N_{\geq 0}$ in $\mathbf{x}$ to $\mathbb{R}^N$ in the $\mathbf{u}$ variables or to the $u'_{N+1} = 1$ hyperplane in $\mathbb{R}^{N+1}$ in the $\mathbf{u}'$ variables.
In this representation each term of the polynomial is proportional to some power of $t$:
\begin{equation}
P(t^\mathbf{u},t) = \sum_{i=1}^{m} c_i \, t^{\langle \mathbf{p}'_i, \mathbf{u}' \rangle}.
\end{equation}
As $t\ll1$, the largest term of the polynomial is the one with the smallest value of $\langle\mathbf{p}'_i, \mathbf{u}'\rangle$, i.e. lowest order in $t$. We can also have several points at the same distance from $\mathbf{u}'$, so several largest terms.

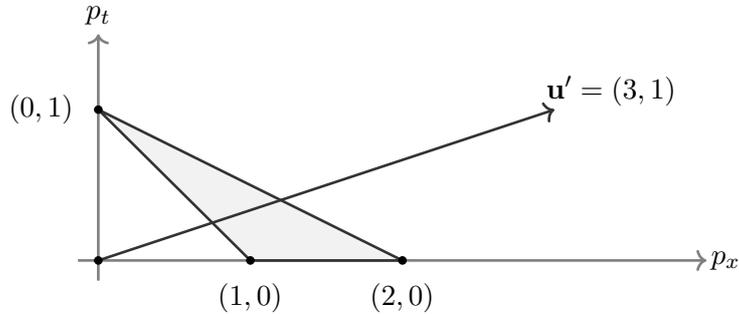
\begin{figure}[h]
    \centering
    {\begin{tikzpicture}
	\begin{pgfonlayer}{nodelayer}
		\node [style=none] (0) at (0, -0.25) {};
		\node [style=none] (1) at (8, 0) {};
		\node [style=none] (2) at (0, 3) {};
		\node [style=dot] (3) at (0, 2) {};
		\node [style=dot] (4) at (2, 0) {};
		\node [style=dot] (5) at (4, 0) {};
		\node [style=none] (10) at (-0.75, 2) {$(0,1)$};
		\node [style=none] (11) at (2, -0.5) {$(1,0)$};
		\node [style=none] (12) at (4, -0.5) {$(2,0)$};
		\node [style=none] (13) at (0, 3.25) {$p_t$};
		\node [style=none] (14) at (8.25, 0) {$p_x$};
		\node [style=none] (23) at (-0.25, 0) {};
		\node [style=none] (24) at (6, 2) {};
		\node [style=dot] (25) at (0, 0) {};
		\node [style=none] (26) at (6.75, 2.25) {$\mathbf{u}'=(3,1)$};
	\end{pgfonlayer}
	\begin{pgfonlayer}{edgelayer}
		\draw [style=line, ->, gray] (0.center) to (2.center);
		\draw [style=line, ->, gray] (23.center) to (1.center);
		\draw [style=line, line join=round, fill={black!5!white}] (3.center)
			 to (4.center)
			 to (5.center)
			 to cycle;
		\draw [style=line, ->] (25) to (24.center);
	\end{pgfonlayer}
\end{tikzpicture}}
    \caption{Newton polytope for $P(x,t) = t+x+x^2$, along with an example vector $\mathbf{u}'$.}
    \label{fig:polytope_simple_uvector}
\end{figure}

We can visualize which terms of the polynomial are larger and which are smaller by looking at the Newton polytope.
Considering vectors $\mathbf{u}'$ in arbitrary directions, subject to the constraint $u'_{N+1}=1$,
we can identify the dominant term as the one with the smallest value of $\langle\mathbf{p}'_i, \mathbf{u}'\rangle$,
i.e. the one where $\mathbf{p}'_i$ projected onto $\mathbf{u}'$ is the smallest.
An example is given in Figure~\ref{fig:polytope_simple_uvector}.

For the polynomial given by \eqref{eq:examplepoly}, 
at the point $\mathbf{u}'=(3,1)$, we have $P(t^\mathbf{u},t) = t+t^3+t^6$, therefore the largest term of the polynomial
\eqref{eq:examplepoly} is $t$, followed by $x$, and the smallest term is $x^2$.

Now we want to assess which expansions defined by a vector $\mathbf{v}$ converge at an integration point defined by $\mathbf{u}'$.
We write the polynomial in the form $P(\mathbf{x}) = Q(\mathbf{x}) + R(\mathbf{x})$, where we have split $P(\mathbf{x})$ into two terms such that $Q(\mathbf{x})$ consists of all the lowest order terms in $t$, and we have suppressed the $t$-dependence in the arguments.
A series expansion in $t$ of $P(\mathbf{x})^m$ is equivalent to a binomial series in $R(\mathbf{x})/Q(\mathbf{x})$ of 
\begin{equation}
P(\mathbf{x})^m = Q(\mathbf{x})^m \left( 1 + \frac{R(\mathbf{x})}{Q(\mathbf{x})} \right)^m.
\end{equation}
This converges for all points $\mathbf{u}'$ where $R(\mathbf{x})/Q(\mathbf{x}) < 1$. From this we conclude that an expansion with a region vector $\mathbf{v}$ converges at a point given by $\mathbf{u}'$ if the lowest order terms along the direction $\mathbf{v}$ contain all the lowest order terms along the direction $\mathbf{u}'$. 
We are then free to assign points in the integration region to different expansion regions, by choosing any region whose lowest order terms in $t$ contain all of the lowest order terms in $t$ of the integration point.

For the example in Fig.~\ref{fig:polytope_simple_uvector}, for $\mathbf{u}'=(3,1)$ we could have an infinite number of different converging expansions, e.g. with a region vector $\mathbf{v}=(1,1)$ or $\mathbf{v}=(10,1)$. However, a good strategy 
is to choose the region vectors to be the vectors normal to the facets of the polytope,
more precisely those normal vectors pointing inwards the polytope and that have a positive $(N+1)$-th component, as we will see in the next section.

\subsubsection{Covering the whole integration domain \label{sec:facets_full_domain}}

In this subsection we will argue that one possible set of expansions covering the whole integration domain is given by the inwards pointing normal vectors to the polytope facets which have a positive $(N+1)$-th component.

First, for any direction $\mathbf{u}$, the vertices with the smallest distance along $\mathbf{u}'$ must be part of some facet of the polytope. Furthermore, the lowest order terms for any $\mathbf{u}'$ must lie on a facet whose inwards pointing normal vector has a positive $(N+1)$-th component, because the vector $\mathbf{u}'$ itself always has positive $(N+1)$-th component.
Therefore, choosing these inwards pointing normal vectors as region vectors, we have a converging expansion for every $\mathbf{u}'$, which means that
with those expansions we have covered the whole integration domain.

\subsubsection{Why multiple expansions vanish \label{sec:multi_exp_vanish}}

When we use expansions of the form described above, the double and higher expansions appearing in Jantzen's general  formula, sketched in \eqref{eq:Jantzen}, vanish.
To see this, let us consider two expansions with region vectors $\mathbf{v}$ and $\mathbf{w}$.
After expanding $P(\{z_1^{v_i}z_2^{w_i}x_i\}, z_1^{v_0}z_2^{w_0}t)$ in $z_1$ and $z_2$, the terms of the produced series are homogeneous in $z$ after introducing any $z$ with $x_i\rightarrow z^{a v_i+bw_i}x_i$, $t\rightarrow z^{a v_0 + bw_0}$, where $a$, $b$ are some constants.
In particular, we can choose $x_i\rightarrow z^{v_i/v_0-w_i/w_0}x_i$, $t\rightarrow t$, which corresponds to only changing the variables $x_i$, and each integral of the series expansion gets a monomial factor in $z$.
These are scaleless integrals, which are zero due to the regularisation we use.
In general, an integral is scaleless whenever upon rescaling the integration variables by some powers of a constant, this constant factorises out of the integrand. For example, in momentum space, this condition corresponds to rescaling a subset $\{k_i\}_b$ of the set of loop momenta $\{k_i\}$ with a parameter $c$ such that
\be
I(\{k_i\}_a,\{c \,k_i\}_b)=c^\alpha I(\{k_i\}) \Rightarrow I(\{k_i\})=0\;,
\ee
where $\alpha$ is the scaling dimension of the integral $I$.
Geometrically this means that the corresponding Newton polytope is not fully dimensional, see also Ref.~\cite{ThesisJensHoff}.

In a similar way, if we do an expansion with a region vector $\mathbf{v}$ that is not a normal vector to a facet, then the result from this region is zero. This is because in that case the smallest order points are on several facets simultaneously, and the series expansion corresponds to expanding successively, using the region vectors of all of those facets.
This multiple expansion is zero by the argument above.

\subsubsection{Why typical Feynman integrals have a commuting region}
\label{sec:commuting}

In Ref.~\cite{Jantzen:2011nz} Jantzen argued that for expansions with non-commuting regions the term $I_{nc}$ vanishes if there is a region which commutes with all other regions.
%For what concerns the non-commuting terms, Jantzen argued that they depend on the boundaries between the regions, which are arbitrary, and therefore the non-commuting terms should be zero.
%The argument should hold whenever there is at least one commuting region. The border shared between one commuting and one non-commuting region can be changed arbitrarily within a certain range.
%However, since only the integrals over the non-commuting region depend on this border, these should not contribute to the final result and should therefore be zero. The argument can then be repeated by considering another border.

With the regions determined using the geometric method, it is easy to see if expansions commute. In general, this happens whenever the facets corres\-ponding to the expansions have a common vertex, otherwise they do not commute.

Using this fact we can argue that whenever we have a polynomial of the typical form $P(\mathbf{x},t) = Q(\mathbf{x}) + t R(\mathbf{x})$, and it has a hard region (i.e. a region vector $(\vec{0},1)$), then the expansion corresponding to the hard region commutes with all the other expansions. This is because all the vertices which are not in the hard region are in the facet with the region vector $(\vec{0},-1)$, which can be ignored due to the negative last component, and therefore all the remaining facets must have at least one vertex from the hard region.

Whenever we have polynomials with positive coefficients of this form, we have at least one expansion commuting with all the others. As discussed previously, this means we can expect the non-commuting expansions in Jantzen's formula to vanish along with the multiple commuting expansions, leaving us with the recipe \eqref{eq:ebrgen} for expansion by regions.

\subsection{Polynomials with coefficients of indefinite sign, threshold expansions}
\label{sec:Glauber}

It is important to note that what we have described works for polynomials with only positive coefficients.
The presence of both positive and negative coefficients can introduce new regions in which the expansions determined through the geometric method would not converge.

In the case of a polynomial with coefficients of different sign, it is possible that for some expansion $Q(\mathbf{x})$ goes to zero and then the assumption $R(\mathbf{x})/Q(\mathbf{x})<1$ does not apply. If the zeros of $Q(\mathbf{x})$ are of order one, we can deform the contour away from the pole of $Q(\mathbf{x})$ and still apply the method of expansion by regions.
However, if $Q(\mathbf{x})$ has a higher order zero, then we need a workaround.
We can for example split the integral into several integrals at the singular point and perform a change of variables~\cite{Ananthanarayan:2018tog,Jantzen:2012mw}.
In general, this means that  we have to check if and how the method of expansion can be applied to
such polynomials.

As a very simple example, we consider the integral
\begin{equation}
\int_0^\infty \frac{dx}{(1+x)^2+t}\;,
\end{equation}
which has only one region $(0,1)$, and has $t^0$ as the lowest order in $t$, but if we change one sign to get
\begin{equation}
\int_0^\infty \frac{dx}{(1-x)^2+t}\;,
\end{equation}
we can no longer use just one region. This is because in this case $Q(x)=(1-x)^2$, which has two zeros at $x=1$ and therefore we cannot deform away from them. We could add a second expansion by treating $(1-x)^2/t$ as small, but it does not commute with the other expansion and therefore we cannot directly apply the method of expansion by regions as in \eqn{eq:ebrgen}.
Instead we may try to split the integral into two integrals at $x=1$ and work with them separately.

Regarding Feynman integrals, there are well-studied cases, like potential and Glauber regions~\cite{Jantzen:2012mw}, where the different signs of the polynomial can effectively cause new regions to arise.
Glauber gluons, which can arise in forward scattering,  are characterised by the fact that they have transverse momenta $k_\perp$ larger than their light-cone components $n\cdot k$ and $\bar{n}\cdot k$ in the Sudakov parametrisation of momenta, 
such that they can induce Coulomb-like interactions among soft and collinear particles~\cite{Becher:2014oda}.
They are intrinsically linked to non-Euclidean kinematics.

As a simple example for a threshold expansion, where different signs present in the graph polynomial cause issues, we consider the one-loop bubble~\cite{Jantzen:2012mw} and use $p^2$ and $y=m^2-p^2/4$ as the parameters, with $p$ being the incoming momentum and $m$ the mass of the particle in the loop (Figure~\ref{fig:1Lbubble}). We are aiming at an expansion in $y$.
The ${\cal F}$-polynomial is then given by
\begin{equation}
{\cal F}=\frac{p^2}{4}(x_1-x_2)^2+y\,(x_1+x_2)^2.
\end{equation}
The geometric method reveals only the hard region. However, in that region, we can have $x_1$ and $x_2$ very close to each other, causing the term $(x_1-x_2)^2$ to be arbitrarily small, but in the hard region the term $y\,(x_1+x_2)^2$ was assumed to be the smallest, meaning this assumption does not always hold in that region. We also cannot use contour deformation to stay away from $x_1=x_2$, because $Q= \frac{p^2}{4}(x_1-x_2)^2$ satisfies the Landau conditions. Thus we cannot apply the method of expansion in the usual way here.
However, we could divide the integration domain into several integrals and perform a change of variables~\cite{Jantzen:2012mw,Ananthanarayan:2018tog,Semenova:2018cwy}, to arrive at different integrals which allow one to directly apply the method of expansion by regions.

\subsection{Additional regulators}

It is well known that expansion by regions introduces spurious singularities that are generally not regulated by dimensional regularisation (DR).
Here we discuss in which cases additional regulators are necessary.
We also derive conditions on the additional regulators guaranteeing that all appearing region integrals are regulated.

In this section it is assumed that the unexpanded integral is fully regulated by DR, though this is not the general case for the class of integrals in \eqref{eq:int}.

The potential singularities of a Feynman integral are reflected by the associated Newton polytope $\Delta$.
It is useful to define the set of $m$-dimensional faces of $\Delta$
\begin{equation}
F^m_{\Delta} = \left\{ f \in \Delta \mid \mathrm{dim}(f) = m \right\}
\end{equation}
and the set of $m$-dimensional faces spanning an affine space that contains the origin
\begin{equation}
F^m_{\Delta,0} =  \left\{ f \in \Delta \mid \mathrm{dim}(f) = m,\; \mathbf{0} \in \mathrm{aff}(f) \right\}.
\end{equation}
The faces of a polytope are again polytopes of lower dimension forming its boundary.
Dimension zero, one, and codimension one faces are called vertices, edges, and facets respectively.
Here $\mathrm{aff}(f)$ is defined as the smallest affine space containing the face $f$.

For each facet $f\in F^{N-1}_{\Delta}$ a singularity is present if the propagator powers $\bm{\nu}$ and the space-time dimension $D$ fulfil the linear inequality
\begin{equation}
\langle \mathbf{n}_f,\bm{\nu}\rangle + a_f \frac{D}{2} \leq 0.
\end{equation}
This condition is derived from the Lee-Pomeransky representation \cite{Lee:2013hzt} using geometric sector decomposition \cite{Kaneko:2009qx,Kaneko:2010kj}, and was given in \cite{Schlenk:2016epj}.
All singularities from facets $f \in F^{N-1}_{\Delta}\setminus F^{N-1}_{\Delta,0}$ are regulated by DR since they have $a_f \neq 0$.
For facets $f \in F^{N-1}_{\Delta,0}$ only normal vectors $\mathbf{n}_f$ with $\langle \mathbf{n}_f, \bm{\nu} \rangle > 0$ for all $\bm{\nu} \in \mathbb{N}^N_{>0}$ are allowed for DR to be sufficient.
This requirement restricts the normal vectors of facets $f \in F^{N-1}_{\Delta,0}$ to $\mathbf{n}_f \in \mathbb{N}^N_{\geq 0}$.
For Feynman integrals the only allowed normal vectors for facets in $F^{N-1}_{\Delta,0}$ are the unit vectors $\hat{\mathbf{e}}_i$ for $1\leq i \leq N$ since $\Delta \in \mathbb{R}^N_{\geq 0}$.

Having discussed the regularisation of the unexpanded integral we can now move on to the regularisation of the region integrals.
As discussed in Section~\ref{sec:geo} the expansion by regions leads to a polyhedral subdivision $\mathrm{SD}(\Delta)$ of the Newton polytope $\Delta$ of the unexpanded integral.
The $N$-dimensional elements of $\mathrm{SD}(\Delta)$ are the Newton polytopes $\Delta_r\in F^N_{\mathrm{SD}(\Delta)}$ of the region integrals $I_r$.
We split the set of $(N-1)$-dimensional facets $F^{N-1}_{\mathrm{SD}(\Delta)}$ into $F^{N-1,\mathrm{ext}}_{\mathrm{SD}(\Delta)}$ consisting of facets that are contained in the boundary of $\Delta$ and the complement $F^{N-1,\mathrm{int}}_{\mathrm{SD}(\Delta)}$ of internal facets that constitute the boundary between different region polytopes $\Delta_r$.

The internal facets can lead to new singularities in those region integrals whose Newton polytopes they are bounding.
Since internal facets with $a_f \neq 0$ and all external facets are regulated by DR, we only have to consider the internal facets $F^{N-1,\mathrm{int}}_{\mathrm{SD}(\Delta),0}$ in the following.
An internal facet $f\in F^{N-1,\mathrm{int}}_{\mathrm{SD}(\Delta),0}$ that forms the boundary between the region polytopes $\Delta_{r_1}$ and $\Delta_{r_2}$ leads to the conditions
\begin{equation}
\langle \mathbf{n}_f, \bm{\nu} \rangle \leq 0
\label{eq:unreg}
\end{equation}
for a singularity in $I_{r_1}$ and
\begin{equation}
\langle \mathbf{n}_f, \bm{\nu} \rangle \geq 0
\end{equation}
for a singularity in $I_{r_2}$.
For any $\bm{\nu}$ at least one of the conditions is fulfilled, which leads to an unregulated singularity.
Additional regulators are therefore required whenever the set $F^{N-1,\mathrm{int}}_{\mathrm{SD}(\Delta),0}$ is non-empty, i.e. $\mathrm{SD}(\Delta)$ contains internal facets with $a_f=0$.

One possibility to regulate the additional singularities is to deform the exponent powers as $\bm{\nu}\rightarrow \bm{\nu} + \delta \bm{\nu}_{\delta}$.
The original integral is recovered in the limit $\delta \rightarrow 0$.

Applying the deformation to \eqref{eq:unreg} leads to the condition
\begin{equation}
\langle \mathbf{n}_f, \bm{\nu}_{\delta} \rangle \neq 0\quad \forall\; f \in F^{N-1,\mathrm{int}}_{\mathrm{SD}(\Delta),0}
\end{equation}
on $\bm{\nu}_{\delta}$ which has to be fulfilled in order for $\delta$ to regulate all additional singularities.
The vector $\bm{\nu}_{\delta}$ has to be chosen such that it lies outside all hyperplanes spanned by internal facets $f$ with $a_f=0$.
It is clear that such a vector can always be found since the $(N-1)$-dimensional hyperplanes spanned by the facets cannot cover the $N$-dimensional space of choices for $\bm{\nu}_{\delta}$ for finite subdivisions $\mathrm{SD}(\Delta)$.

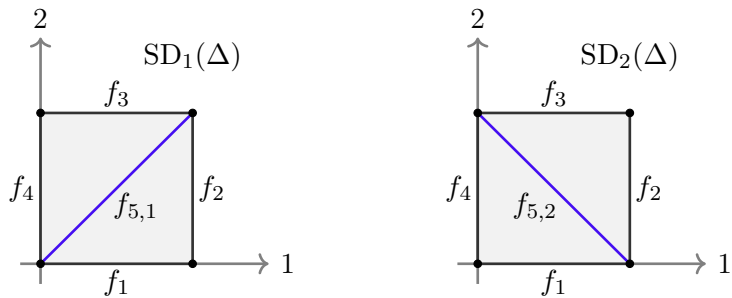
\begin{figure}
    \centering
    {\begin{tikzpicture}
	\begin{pgfonlayer}{nodelayer}
		\node [style=dot] (0) at (0, 0) {};
		\node [style=dot] (1) at (2, 0) {};
		\node [style=none] (2) at (3, 0) {};
		\node [style=dot] (3) at (0, 2) {};
		\node [style=none] (4) at (0, 3) {};
		\node [style=dot] (5) at (2, 2) {};
		\node [style=none] (6) at (-0.25, 0) {};
		\node [style=none] (7) at (0, -0.25) {};
		\node [style=none] (8) at (3.25, 0) {1};
		\node [style=none] (9) at (0, 3.25) {2};
		\node [style=none] (10) at (1, -0.25) {$f_1$};
		\node [style=none] (11) at (2.25, 1) {$f_2$};
		\node [style=none] (12) at (1, 2.25) {$f_3$};
		\node [style=none] (13) at (-0.25, 1) {$f_4$};
		\node [style=none] (14) at (1.25, 0.75) {$f_{5,1}$};
		\node [style=dot] (15) at (5.75, 0) {};
		\node [style=dot] (16) at (7.75, 0) {};
		\node [style=none] (17) at (8.75, 0) {};
		\node [style=dot] (18) at (5.75, 2) {};
		\node [style=none] (19) at (5.75, 3) {};
		\node [style=dot] (20) at (7.75, 2) {};
		\node [style=none] (21) at (5.5, 0) {};
		\node [style=none] (22) at (5.75, -0.25) {};
		\node [style=none] (23) at (9, 0) {1};
		\node [style=none] (24) at (5.75, 3.25) {2};
		\node [style=none] (25) at (6.75, -0.25) {$f_1$};
		\node [style=none] (26) at (8, 1) {$f_2$};
		\node [style=none] (27) at (6.75, 2.25) {$f_3$};
		\node [style=none] (28) at (5.5, 1) {$f_4$};
		\node [style=none] (29) at (6.5, 0.75) {$f_{5,2}$};
		\node [style=none] (30) at (2, 2.75) {$\mathrm{SD}_1(\Delta)$};
		\node [style=none] (31) at (7.75, 2.75) {$\mathrm{SD}_2(\Delta)$};
	\end{pgfonlayer}
	\begin{pgfonlayer}{edgelayer}
		\draw [style=line, ->, gray] (6.center) to (2.center);
		\draw [style=line, ->, gray] (7.center) to (4.center);
		\draw [style=line, fill={black!5!white}] (3.center)
			 to (5.center)
			 to (1.center)
			 to (0.center)
			 to cycle;
		\draw [style=line, draw={rgb,255: red,69; green,14; blue,241}] (0) to (5);
		\draw [style=line, ->, gray] (21.center) to (17.center);
		\draw [style=line, ->, gray] (22.center) to (19.center);
		\draw [style=line, fill={black!5!white}] (18.center)
			 to (20.center)
			 to (16.center)
			 to (15.center)
			 to cycle;
		\draw [style=line, draw={rgb,255: red,69; green,14; blue,241}] (18) to (16);
	\end{pgfonlayer}
\end{tikzpicture}}
    \caption{Example of subdivisions generated by expansion by region of integrals as defined in \eqref{eq:int} with $P_1 = 1 + t x_1 + x_1 x_2 + t x_2 $ (left) and $P_2 = t +  x_1 + t x_1 x_2 +  x_2 $ (right) in the limit $t\rightarrow0$.}
    \label{fig:subdivex}
\end{figure}
As an example Fig.~\ref{fig:subdivex} shows two subdivisions $\mathrm{SD}_1(\Delta)$ and $\mathrm{SD}_2(\Delta)$ appearing in the expansion of integrals $I_1$ and $I_2$ with $P_1 = 1 + t x_1 + x_1 x_2 + t x_2 $ and $P_2 = t +  x_1 + t x_1 x_2 +  x_2 $ in the limit $t\rightarrow 0$.
Even though the integrals are not Feynman integrals, the set of external facets $F^{N-1,\mathrm{ext}}_{\mathrm{SD}_1(\Delta)} = F^{N-1,\mathrm{ext}}_{\mathrm{SD}_1(\Delta)} = \left\{ f_1, f_2, f_3, f_4 \right\}$ is regulated by DR.
For the subdivision $\mathrm{SD}_1(\Delta)$ one has $F^{N-1,\mathrm{int}}_{\mathrm{SD}_1(\Delta),0}  = \left\{ f_{5,1} \right\}$, therefore an additional regulator with the condition $\langle \mathbf{n}_{f_{5,1}},\bm{\nu}_{\delta}\rangle = (\bm{\nu}_{\delta})_1 - (\bm{\nu}_{\delta})_2 \neq 0$ on the components of $\bm{\nu}_{\delta}$ is necessary.
For $\mathrm{SD}_2(\Delta)$ the set $F^{N-1,\mathrm{int}}_{\mathrm{SD}_2(\Delta),0}$ is empty since the line spanned by the internal facet $f_{5,2}$ does not contain the origin and no additional regulator is required.

\section{New features of \pysecdec}
\label{sec:newfeatures}

In this section, we list and briefly introduce the new features of \pysecdec{}, which, apart from the option to perform asymptotic expansions, represent a major step in the ability of the code to evaluate \textit{amplitudes} rather than individual integrals.
The following changes have been made compared to \pysecdec{} version 1.4.5:
\begin{itemize}
\item Asymptotic expansions can be performed in an automated way.

\item The combination of several integrals together with their corresponding coefficients can be evaluated as a sum weighted by its numerical importance,
i.e.  each term is evaluated with the appropriate number of samplings such that a global accuracy goal for the sum is reached most efficiently.

\item The contour deformation parameters $\lambda_i$ are now adjusted automatically, so that the integration will no longer stop with the ``\textit{sign check error}'' if the original $\lambda_i$ values lead to an invalid integration contour.

\item The \texttt{WorkSpace} parameter of \form{}~\cite{Kuipers:2013pba,Ruijl:2017dtg} is now automatically increased if \form{} fails due to insufficient \texttt{WorkSpace}. Users are no longer required to adjust the \mintinline{python}{form_work_space} parameter of \mintinline{python}{make_package} and \mintinline{python}{loop_package}.

\item The installation procedure was simplified, and \pysecdec{} can now be installed from the Python Package Index\footnote{\url{https://pypi.org/project/pySecDec/}} using
\begin{code}
\begin{minted}
{shell}
python3 -m pip install --user pySecDec
\end{minted}
\end{code}

\item The functions \mintinline{python}{series_to_ginac}, \mintinline{python}{series_to_sympy}, \mintinline{python}{series_to_maple} and
\mintinline{python}{series_to_mathematica} have been added. They convert the output of \pysecdec{} to a syntax more suitable for use with various computer algebra systems. 

\item The support for Python version~2.7 was dropped in favour of version~3.6 or newer. Python is now always invoked as \mintinline{shell}{python3}.

\end{itemize}

The following new functions have been introduced:
\begin{itemize}
\item \texttt{make\_regions} provides a package generator for performing expansion by regions of general parameter integrals.
\item \texttt{loop\_regions} a wrapper around \texttt{make\_regions} which simplifies applying expansion by regions to loop integrals.
\item \texttt{sum\_package} generates a C++ library for computing the sum of (weighted) integrals/sectors.
\end{itemize}
They are documented in detail in the online documentation\footnote{\url{https://secdec.readthedocs.io}} distributed with the code and examples of their usage are shown in Section~\ref{sec:usage} and Section~\ref{sec:examples}.

The functions \texttt{make\_package} and \texttt{loop\_package} now also generate libraries that integrate individual sectors separately and sum them with unit coefficient, this allows the code to optimise how precisely each sector is calculated in order to reach the user's requested precision in the shortest time possible.

\section{Using the new features of \pysecdec}
\label{sec:usage}

%\subsection{Usage}

\subsection{Usage of \mintinline{python}{sum_package}}

\begin{listing}[th!]
\begin{minted}
[
frame=lines,
framesep=2mm,
baselinestretch=1.2,
fontsize=\footnotesize,
linenos
]
{python}
from pySecDec import Coefficient, MakePackage, sum_package

if __name__ == "__main__":

    common_args = {}
    common_args['real_parameters'] = ['s']
    common_args['regulators'] = ['eps']
    common_args['requested_orders'] = [0]

    coefficients = [
        [ # sum1
            Coefficient(['2*s'],['1'],['s']),   # easy1
            Coefficient(['3*s'],['1'],['s'])    # easy2
        ],
        [ # sum2
            Coefficient(['s'],['2*eps'],['s']), # easy1
            Coefficient(['s*eps'],['3'],['s'])  # easy2
        ]
    ]

    integrals = [
        MakePackage('easy1',
            integration_variables = ['x','y'],
            polynomials_to_decompose = ['(x+y)^(-2+eps)'],
            **common_args),
        MakePackage('easy2',
            integration_variables = ['x','y'],
            polynomials_to_decompose = ['(2*x+3*y)^(-1+eps)'],
            **common_args)
    ]
    
    # generate code sum of (int * coeff)
    sum_package('easy_sum', integrals,
        coefficients = coefficients, **common_args)
\end{minted}
\caption{Python script for generating a code which evaluates a sum of weighted integrals. The complete example code can be found in the \mintinline{python}{examples/easy_sum} folder.}
\label{code:sum}
\end{listing}

The new \mintinline{python}{sum_package} function generates a library which efficiently evaluates weighted sums of integrals.
The function expects a list of package generators and an array of coefficients.
The supported package generators are \mintinline{python}{MakePackage} and \mintinline{python}{LoopPackage}, which expect the same arguments as  \mintinline{python}{make_package} and  \mintinline{python}{loop_package}, respectively.
The \mintinline{python}{coefficients} array is an $M \times N$ list of lists, where $M$ is the number of amplitudes (or weighted sums) to be evaluated and $N$ is the number of integrals in each sum.
The generated library will attempt to obtain the user's requested precision for each weighted sum in the shortest time possible, results for each integral are shared between the different weighted sums.

In Code Example~\ref{code:sum} we show a simple usage of \mintinline{python}{sum_package}.
The code will generate a library which simultaneously evaluates the following two sums,
\begin{eqnarray}
& 2 s\ I_1 + 3 s\ I_2, \nonumber \\
& \frac{s}{2 \epsilon}\ I_1 + \frac{s \epsilon}{3}\ I_2, \nonumber
\end{eqnarray}
where $I_1$ and $I_2$ are the integrals \mintinline{python}{easy1} and \mintinline{python}{easy2}, $\epsilon$ is the regulator and $s$ is a real parameter specified at run time.
The \mintinline{python}{Coefficient} class represents a rational function; it expects a list of numerators, a list of denominators and a list of real or complex parameters present in the coefficient. 
The terms in the numerator (or denominator) are multiplied together after evaluation, this allows the user to avoid expanding factorised expressions.
Coefficients with poles in the regulator are supported, the \mintinline{python}{sum_package} function will determine the leading power in each regulator and ensure that the integrals they multiply are expanded to a sufficient order in each regulator to reach the \mintinline{python}{requested_orders}.

\subsection{Usage of the expansion options}

In \pysecdec{} both rescaling methods, the $z$-method \eqref{eq:z-method} and
the $t$-method \eqref{eq:t-method}, can be used. The examples provided with the code
demonstrate both methods.
More details about the two methods and their numerical behaviour are
given in Appendix~\ref{sec:appendix1}.

Code Examples~\ref{code:1Ldotbubble} and~\ref{code:1Ldotbubble-int} show how to generate and integrate a
one-loop two-point function with one propagator squared---an example
already discussed in Section \ref{sec:momspace} and shown in Fig.~\ref{fig:dotted_bubble}.
The corresponding \pysecdec{} files can be found in
the \texttt{bubble1L\_dotted\_ebr} subfolder of the \texttt{examples}
folder.

As discussed above, the expansion in $m^2$ can be performed in two
ways: (1)~by replacing $m^2$ with $zm^2$,
expanding in $z$, and eventually setting $z=1$; or (2)~by expanding directly in $m^2$.
Both methods are demonstrated by generation and integration files in
the folder {\tt examples/bubble1L\_dotted\_ebr}; here we focus on (2),
the $t$-method.

In line~7, the corresponding integral is defined.
One can see from line~11 that the field \mintinline{python}{regulators}
is a list, which can contain more regulators than just the dimensional
regulator $\eps$.

In line~16, the function \mintinline{python}{loop_regions} is called.
In line~19 the expansion parameter $msq$ is specified.
In line~24, \mintinline{python}{sum_package} is called, which
generates code for each region and for each order up to (and including) \mintinline{python}{expansion_by_regions_order},
as well as code to evaluate the sum
over all regions up to the requested order.

If we instead wished to use the $z$-method, we would specify
\mintinline{python}{real_parameters} \mintinline{python}{=[psq,msq,z]} in line~27. In the integration file (Code Example~\ref{code:1Ldotbubble-int}) we would also set $z=1$ in line~8.
For both the $t$-method and the $z$-method, \mintinline{python}{msq} is set to the desired (small) value.

Expanding up to order one in the smallness parameter, the numerical result is very close to the full result. 

\begin{listing}[th!]
\begin{minted}
[
frame=lines,
framesep=2mm,
baselinestretch=1.2,
fontsize=\footnotesize,
linenos
]
{python}
from pySecDec import LoopIntegralFromPropagators, loop_regions, sum_package

# guard the code to allow multiprocessing inside pySecDec
if __name__ == "__main__":

    # define the loop integral for the case where we expand in msq
    li_m = LoopIntegralFromPropagators(
        propagators = ("(k+p)**2", "k**2-msq"),
        loop_momenta = ["k"],
        powerlist = [1, 2],
        regulators = ["eps"],
        replacement_rules = [("p*p", "psq")])

    # find the regions and expand the integrals using expansion
    # by regions
    sum_terms = loop_regions(
        name = "bubble1L_dotted_m",
        loop_integral = li_m,
        smallness_parameter = "msq",
        expansion_by_regions_order = 1)

    # generate code that will calculate the sum of all regions
    # and the requested orders in the smallness parameter
    sum_package("bubble1L_dotted_m", sum_terms,
        regulators = li_m.regulators,
        requested_orders = [0],
        real_parameters = ["psq", "msq"],
        complex_parameters = [])
\end{minted}
\caption{Python script to generate the code that uses the $t$-method expansion
  to calculate the dotted bubble example in
  Figure~\ref{fig:dotted_bubble}.}
\label{code:1Ldotbubble}
\end{listing}

\begin{listing}[th!]
\begin{minted}
[
frame=lines,
framesep=2mm,
baselinestretch=1.2,
fontsize=\footnotesize,
linenos
]
{python}
from pySecDec.integral_interface import IntegralLibrary, series_to_sympy
import sympy as sp

if __name__ == "__main__":

    psq, msq = 4, 0.002
    name = "bubble1L_dotted_m"
    real_parameters = [psq, msq]

    # load the library
    intlib = IntegralLibrary(f"{name}/{name}_pylink.so")
    intlib.use_Qmc(transform="korobov3")

    # integrate
    integral_without_prefactor, prefactor, integral_with_prefactor = \
        intlib(real_parameters)

    # convert the result to sympy expressions
    result, error = \
        map(sp.sympify, series_to_sympy(integral_with_prefactor))

    # access and print individual terms of the expansion
    print("Numerical Result")
    for power in [-2, -1, 0]:
        val = complex(result.coeff("eps", power))
        err = complex(error.coeff("eps", power))
        print(f"eps^{power:<2} {val: .5f} +/- {err:.5e}")
\end{minted}
    \caption{Python script to perform the integration using the library generated in Code Example~\ref{code:1Ldotbubble}. As usual, the library must be built by running \mintinline{shell}{make -C bubble1L_dotted_m} before it can be used.}
    \label{code:1Ldotbubble-int}
\end{listing}

It is interesting to note that for the point $p^{2} = 1, m^{2} = 0.1$,
only the $z$-method converges in \pysecdec.
This is because the methods are related by a change of variables,
which moves the location of the poles of the integrand.
For these parameter values in combination with the $t$-method, sector
decomposition will divide the integration domain into sectors such
that the boundary coincides with a pole.
However, as the contour deformation vanishes at the endpoints, a pole at the endpoint of a sector results in an ill-defined integral.
Therefore we get a result only with the $z$-method, for this
particular point, which is  
\begin{align}
I&=(+2.2 {\cdot} 10^{-16} \pm 3.6 {\cdot} 10^{-16}) \, \eps^{-1} + \nn\\
 &+(+2.20259 \pm 7.2 {\cdot} 10^{-16} + (-3.14159 \pm 1.5 {\cdot} 10^{-15}) \,i) \, + \nn\\
 &+\mathcal{O}(\eps)\;.
\end{align}
A more detailed discussion of this issue is given in Appendix~\ref{subsec:convergence_two_methods}.

We should emphasize that the $t$-method in general shows much better convergence and therefore should preferably be used.
Only for cases as the one described above the $z$-method is clearly the better choice.

\subsection{Comparison to standard \pysecdec{} and timings}
\label{sec:timings}

\begin{figure}
\includegraphics[width=1.\textwidth]{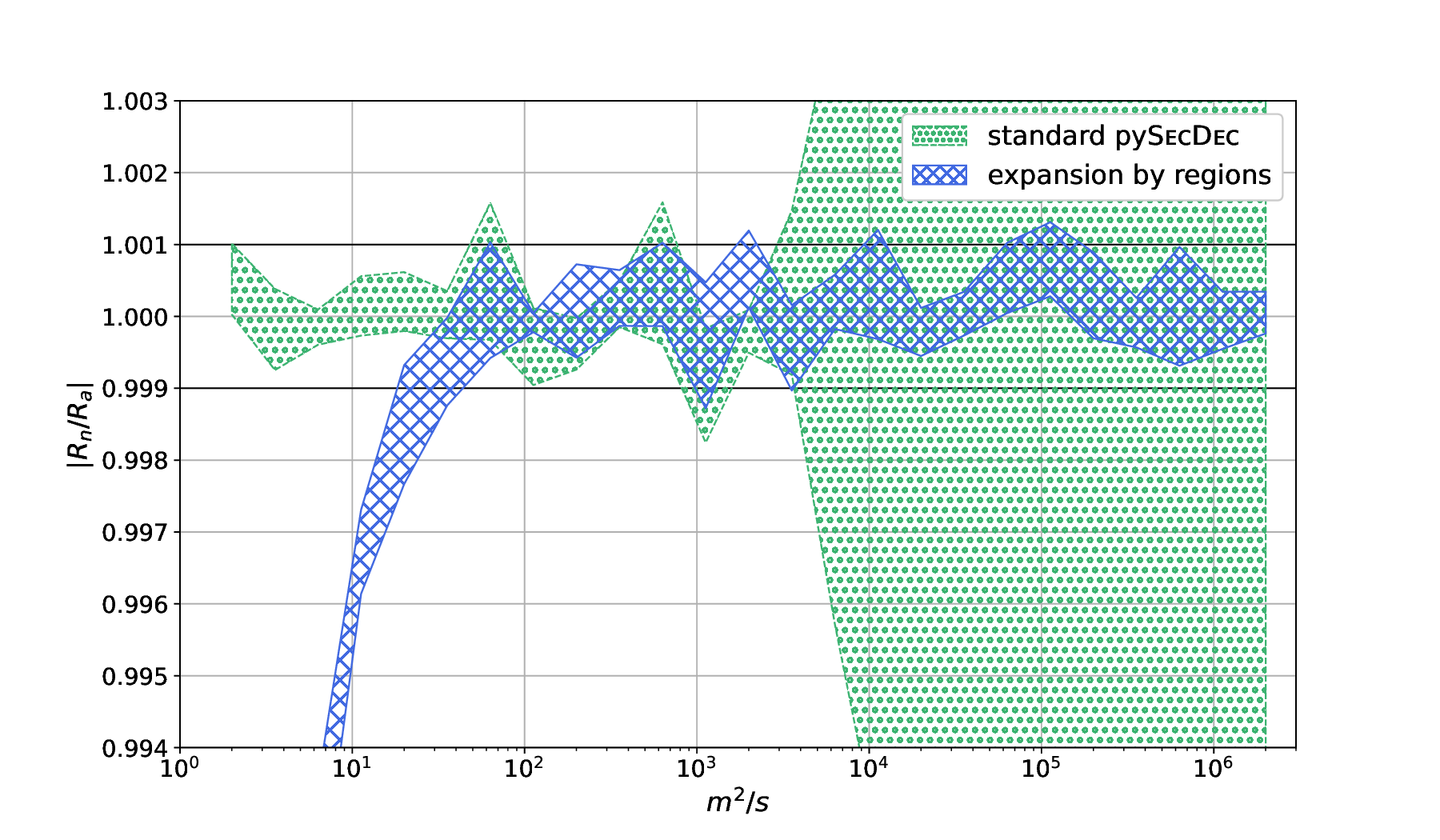} 
\caption{Scan over different orders of magnitude of $r = m^2/s$ for the first of the three two-loop triangles given in Table~\ref{tab:timings}. The modulus of the ratio between the numerical result and the analytic result, $|R_{n}/R_{a}|$, is plotted against $r$. The bands indicate the numerical uncertainties on the result.}
\label{fig:ScanAnalytic}
\end{figure}

In this section we compare expansion by regions with standard \pysecdec{}. We evaluate the three two-loop triangles given in Table~\ref{tab:timings} at order 0 in the $\eps$-expansion for different kinematical configurations. The three integrals are characterised by the two kinematic parameters $s$ and $m^{2}$. We perform the integration with standard \pysecdec{}, setting the ratio $r$ of kinematic scales first to $10^{1}$ and then to $10^{3}$, while for expansion by regions we evaluate them just for $r=10^{3}$.
The limits we consider for the expansion are $s \ll m^{2}$ for the 1st and 3rd triangles, $m^{2} \ll s$ for the 2nd one. \mintinline{python}{expansion_by_regions_order} is set to~0 for all integrals. Note that this expansion order is the leading order for the last two triangles, while it is up to subleading order for the first.
The accuracy goal\footnote{The integration stops whenever the accuracy goal or the maximum number of evaluations are reached, meaning that the actual final error can be smaller than the accuracy goal.}, i.e. the relative error to be achieved on the result, is fixed to $10^{-4}$. Note that here we consider only the error due to the numerical integration, not due to the expansion. The latter is taken into account later in the discussion of Fig.~\ref{fig:ScanAnalytic}. The integration times for both \pysecdec{} and expansion by regions are given in Table~\ref{tab:timings}.
As one can see, for $r = 10^{1}$, \pysecdec{} reaches the given accuracy in less than a minute, but for the first two of the integrals the integration times increase considerably when we increase the scale ratio to $r = 10^{3}$. Expansion by regions instead reaches the demanded numerical accuracy in a shorter amount of time. This behaviour is expected since the expansion removes the huge scale differences among the terms in the integrands (at least with the $t$-method, with the $z$-method large scale differences are removed only for the hard region), which can lead to numerical instabilities. Considering the third example in Table~\ref{tab:timings} without expansion by regions, after sector decomposition the constant term of the $\mathcal{F}$-polynomial is always $m^2$ and never the small parameter $s$. This means that we do not expect any numerical issues for small values of $s$ (or even $s=0$), which is confirmed by the results.
Expansion by regions for this integral leads to only one region, the hard region, in which case the $z$-method and the $t$-method produce the same integrals, hence no difference between them is expected.

We should mention however that the convergence properties can depend critically on the actual scale that is used as the smallness parameter. In particular, with the $t$-method, the first and second integrals in Table~\ref{tab:timings} do not converge whenever we set $m^2=1$ and $s=1$, respectively, creating an unregulated spurious singularity at the upper integration limit,
as explained in Appendix~\ref{subsec:convergence_two_methods}.
However, they do converge very well for $s=0.004, m^2=4$ ($m^2=0.004, s=4$), which are the values used for Table~\ref{tab:timings}.

\begin{figure}
\includegraphics[width=1.\textwidth]{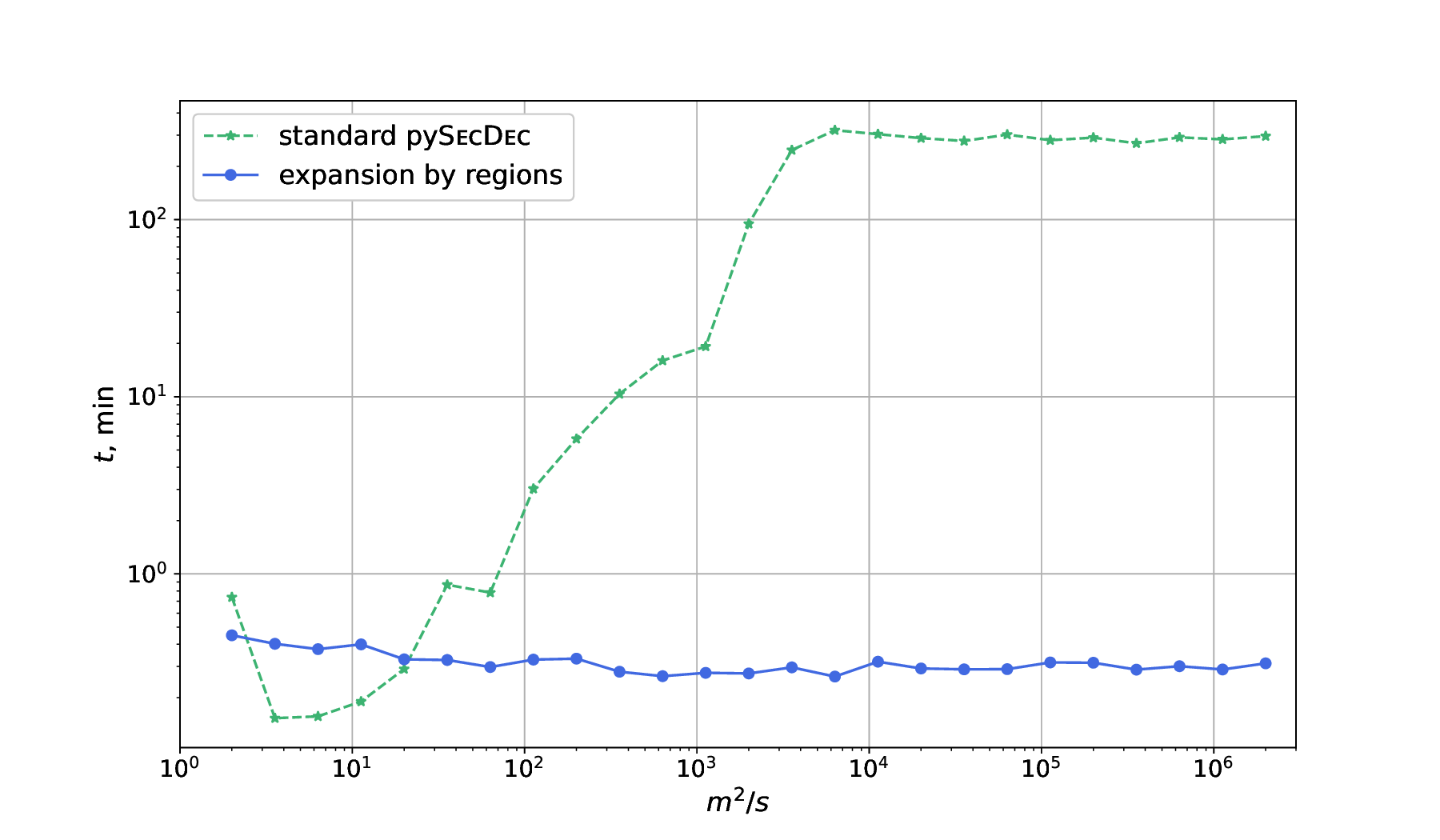} 
    \caption{Scan over different orders of magnitude of $r = m^2/s$ for the first of the three two-loop triangles given in Table~\ref{tab:timings}. Integration times (in minutes) are plotted against $r$. The relative accuracy goal is $10^{-3}$; the wall clock limit has been set to 5~hours. The dashed lines denote the evaluation with standard \pysecdec, the solid lines denote the evaluation with the expansion by regions option of \pysecdec.}
\label{fig:Scan}
\end{figure}

A more detailed analysis is shown in Figs.~\ref{fig:Scan} and~\ref{fig:ScanAnalytic} for the first of the three two-loop triangles in Table~\ref{tab:timings}.  Fig.~\ref{fig:Scan} shows the integration times for both expansion by regions (solid line) and \pysecdec{} (dashed line) plotted against the ratio of the invariants.  Fig.~\ref{fig:ScanAnalytic} shows $|R_{n}/R_{a}|$, the modulus of the ratio of the finite parts between the numerical result obtained with expansion by regions and \pysecdec{} and the analytic result (from Ref.~\cite{Fleischer:1998nb}), plotted against the ratio of the invariants. The shaded areas are given by adding and subtracting the numerical uncertainty to the result. The accuracy goal is fixed to $10^{-3}$. From Fig.~\ref{fig:Scan} one can see how the integration times for \pysecdec{} blow up as the ratio of scales increases, while expansion by regions is numerically stable over the different orders of magnitude.

Fig.~\ref{fig:ScanAnalytic} shows that the error due to the expansion is no longer dominant compared to the integration error if $m^2/s \gtrsim 40$, and that the results from \pysecdec{} and expansion by regions are compatible with each other in this case. Combining the information of the two plots, it is clear that whenever the approximation due to the expansion is negligible compared to the numerical uncertainty of the result, using the expansion by regions option instead of standard \pysecdec{} is the better choice to evaluate the integrals in the given kinematic limit. As a final remark, it is worth to point out that the above considerations for some integrals might hold only much deeper in the kinematic limit.

\begin{table}[htb]
\centering
\begin{tabular}{*{5}{M{2.0cm}}} \\
\toprule 
Diagram & \texttt{psd}  [s] \quad $r=10^{1}$& \texttt{psd} [s] \quad $r=10^{3}$ & \texttt{ebr\_t} [s] $r=10^{3}$  & \texttt{ebr\_z} [s] $r=10^{3}$\\
\midrule
\sfig{plots/triangle2L.tikz}     & 31 & 9900 & 14 & 41 \\
\midrule
\sfig{plots/triangle2L_GH2.tikz} & 34 & 38000 & 70 & 350 \\
\midrule
\sfig{plots/triangle2L_GH1.tikz} & 2.6 & 2.6 & 1.1 & 0.9 \\
\bottomrule \\
\end{tabular}
\caption{Comparison of the timings (in seconds) between standard \pysecdec{} (\texttt{psd}) and expansion by regions with the $t$-method (\texttt{ebr\_t}) and the $z$-method (\texttt{ebr\_z}) to reach a relative accuracy of $10^{-4}$, for the ratio of scales denoted by $r$. The timings refer to the integration time, not including code generation and compilation. They have been obtained on an {\sc Nvidia} Tesla A100 GPU (plus 2 CPU threads, mainly for service), using the {\sc Qmc} integrator.}\label{tab:timings}
\end{table}

\section{Examples}
\label{sec:examples}

The following examples should demonstrate the usage of expansion by
regions within \pysecdec{} and can be found in the folder {\tt examples}.
For each integral, the prefactor is the one specified in
Eq.~(\ref{eq:feynint}) unless a reference to the literature is
given. In this case the prefactor is the one used in the reference.

Each example with integral name \texttt{name} was carried out by first running the script \texttt{generate\_name.py},
then compiling it with \texttt{make -C name}, then finally running the script \texttt{integrate\_name.py}.

For the numerical values we have used the \qmc{} integrator with
\mintinline{python}{transform} set to \mintinline{python}{'korobov3'}.
The accuracy goal was set to $\eps_{\rm{rel}}=10^{-2}$ unless stated otherwise.

\subsection{One-loop bubble}

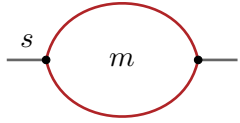
\begin{figure}[htb]
  \centering
  {\begin{tikzpicture}
	\begin{pgfonlayer}{nodelayer}
		\node [style=none] (0) at (-1.5, 0) {};
		\node [style=dot] (1) at (-1, 0) {};
		\node [style=dot] (2) at (1, 0) {};
		\node [style=none] (3) at (1.5, 0) {};
		\node [style=none] (4) at (0, 0) {$m$};
		\node [style=none] (5) at (-1.25, 0.25) {$s$};
	\end{pgfonlayer}
	\begin{pgfonlayer}{edgelayer}
		\draw [style=edge, bend left=75, looseness=1.25] (1) to (2);
		\draw [style=external edge] (0.center) to (1);
		\draw [style=external edge] (2) to (3.center);
		\draw [style=edge, bend right=75, looseness=1.25] (1) to (2);
	\end{pgfonlayer}
\end{tikzpicture}}
  \caption{One loop bubble integral with equal internal masses.}
  \label{fig:1Lbubble}
\end{figure}

The example \texttt{bubble1L\_ebr} calculates a one-loop bubble integral with equal internal masses, see Fig.~\ref{fig:1Lbubble}. Its purpose is to show the basic usage of expansion by regions within \pysecdec{}. The kinematics invariant $s$ and the internal mass squared $m^{2}$ can be set in \texttt{real\_parameters} in the integration file.

We consider as separate cases the large mass expansion ($m^2 \gg s$) using the smallness parameter $s$ and the small mass expansion ($m^2 \ll s$) using the smallness parameter $m^2$. 

\medskip

In the large mass limit, there is only one region with region vector $(0,0,1)$.
The result, which starts at order 0 and is expanded up to order 2 in the smallness parameter, for the point $s=0.002$, $m^{2} = 4$ reads
\begin{align}
I&=(+1.00000 \pm 2.7 {\cdot} 10^{-11} + (-1.0 {\cdot} 10^{-11}\pm 7.0 {\cdot} 10^{-11} ) \, i) \, \eps^{-1}+ \nn\\
 &+(-1.96344 \pm 3.1 {\cdot} 10^{-11} + (+3.23330 {\cdot} 10^{-5} \pm 8.0 {\cdot} 10^{-11} ) \, i) \, + \nn\\
 &+\mathcal{O}\left(\eps, \left(s/m^2\right)^3\right).
\end{align}

\medskip

In the small mass limit, there are 3 regions with region vectors $(-1,0,1)$, $(0,-1,1)$, and $(0,0,1)$. The result, which starts at order 0 and is expanded up to order 2 in the smallness parameter, for the point $s=4$, $m^{2} = 0.002$ reads
\begin{align}
I&=(+1.00000 \pm 2.4 {\cdot} 10^{-15}                 + (-1.72 {\cdot} 10^{-14}\pm 9.8 {\cdot} 10^{-15} ) \, i) \, \eps^{-1}+ \nn\\
 &+(+4.50944 {\cdot} 10^{-2} \pm 1.3 {\cdot} 10^{-14} + (+3.13844 \pm 1.5 {\cdot} 10^{-14} ) \,i) \, + \nn\\
 &+\mathcal{O}\left(\eps,  \left(m^2/s\right)^3\right).
\end{align}

\subsection{One-loop box for $gg\to HH$ in the high-energy limit}

\begin{figure}[htb]
   \centering
   {\begin{tikzpicture}
	\begin{pgfonlayer}{nodelayer}
		\node [style=none] (0) at (-1.25, 1.25) {};
		\node [style=none] (1) at (1.25, 1.25) {};
		\node [style=none] (2) at (1.25, -1.25) {};
		\node [style=none] (3) at (-1.25, -1.25) {};
		\node [style=dot] (4) at (-0.75, 0.75) {};
		\node [style=dot] (5) at (0.75, 0.75) {};
		\node [style=dot] (6) at (0.75, -0.75) {};
		\node [style=dot] (7) at (-0.75, -0.75) {};
		\node [style=none] (8) at (0, 0) {$m_t$};
		\node [style=none] (9) at (1.5, 0.75) {$m_H$};
		\node [style=none] (10) at (1.5, -0.75) {$m_H$};
	\end{pgfonlayer}
	\begin{pgfonlayer}{edgelayer}
		\draw [style=edge] (4) to (5);
		\draw [style=edge] (5) to (6);
		\draw [style=edge] (6) to (7);
		\draw [style=edge] (7) to (4);
		\draw [style=external edge] (1.center) to (5);
		\draw [style=external edge] (2.center) to (6);
		\draw [style=massless external edge] (3.center) to (7);
		\draw [style=massless external edge] (0.center) to (4);
	\end{pgfonlayer}
\end{tikzpicture}}
   \caption{One-loop box integral with a massive  quark loop and two massive legs.}
   \label{fig:boxHH}
 \end{figure}
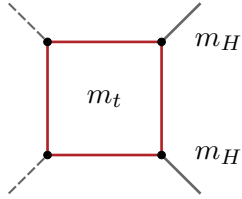

 The example \texttt{box1L\_ebr} demonstrates a high-energy
 expansion of a multi-scale process with massive propagators and
 massive external legs, as it occurs for example in the process $gg\to HH$, see also Ref.~\cite{Mishima:2018olh}.
 The corresponding one-loop box integral, shown in Fig.~\ref{fig:boxHH}, is given by
\begin{align}
&  I_4(\delta_1,\delta_2,\delta_3,\delta_4)=\int_{-\infty}^{\infty}\frac{d^Dk}{i\pi^{\frac{D}{2}}}\;\times \\
&\frac{1}{[k^2-m_t^2]^{\delta_1}[(k+p_1)^2-m_t^2]^{\delta_2}[(k+p_1+p_2)^2-m_t^2]^{\delta_3}[(k-p_4)^2-m_t^2]^{\delta_4}}\nn\;,
\end{align}
where the Feynman $i\eps$-prescription is implicit.
The corresponding \pysecdec{} files show the calculation of 
the box integral in the high energy
limit $m_H^2< m_t^2\ll s, |t|, |u|$, up to lowest order in both $m_H^2$ and $m_t^2$.

For the genuine one-loop case, the propagator powers would be $\delta_i=1\,\forall i$, however we need to introduce analytic regulators because dimensional regularisation will not regulate some of the intermediate spurious poles occurring in the expansions.
In Ref.~\cite{Mishima:2018olh}, a different regulator is chosen for each propagator, i.e. $\delta_i=1+n_i\, , i=1,2,3,4$.
However, to avoid the proliferation of expansions in different regulators, even though this would be technically possible in \pysecdec,
it is here more convenient (and sufficient) to choose one extra regulator and rescale it by rational numbers which have no common
divisor for each propagator, such as $\delta_1=1+n_1/2, \delta_2=1+n_1/3, \delta_3=1+n_1/5, \delta_4=1+n_1/7$.
Note that the regulator $n_1$ is given first in the list of regulators.
The order is important in this list, as the expansion in $\eps$ and in the extra regulators do in general not commute.
The original integral is only independent of the analytic regulator for non-zero $\eps$.
Therefore, in order to have a result independent of the analytic regulator, one should expand in the analytic regulator first and cancel the spurious poles before expanding in $\eps$.
\pysecdec{} will do the expansion in the same order as given in the list.
The expansion order in each of the regulators is given in the field \mintinline{python}{requested_orders}.
The setting \mintinline{python}{[0,0]} means that the expansion in the regulators is performed up to order $n_1^0$ and $\eps^0$.

Expanding in $m_H^2$ there is only one region vector $(0,0,0,0,1)$, so to get the lowest order term we can set $m_H^2$ to zero.
Our smallness parameter is then given by $m_t^2$, and we only retain the terms up to order $(m_t^2)^0$.
This leads to five regions, four of which are related due to the symmetries of the diagram.
Note however that the analytic regulators need to be permuted to make use of these symmetries. 
The corresponding region vectors are
$(0,0,0,0,1)$, $(-1,-1,0,0,1)$, $(0,0,-1,-1,1)$, $(-1,0,0,-1,1)$ and $(0,-1,-1,0,1)$.
Note that the last element of the region vector does not refer to the scaling of a Feynman parameter,
but to the exponent of the smallness parameter itself, and that shifts of all but the last component of the region vector by the same constant do not change the result.
Therefore, shifting for example with $(1,1,1,1,0)$ would lead to region vectors with positive entries.

The lowest order result in the smallness parameter for the point $s = 4.0, t = -2.82843, m^{2}_{t} = 0.1$  reads
\begin{align}
I&=-1.30718 \pm 2.7 {\cdot} 10^{-6} + (1.85618 \pm 3.0 {\cdot} 10^{-6} ) \, i + \mathcal{O}\left(\eps,n_1, \frac{m_t^2}{s},\frac{m_t^2}{t}\right),
\end{align}
in agreement with the expected value.

\subsection{One-loop triangles}

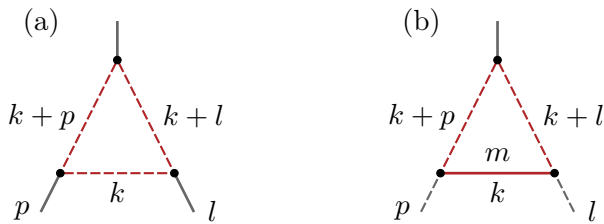
\begin{figure}[htb]
  \centering
  {\begin{tikzpicture}
	\begin{pgfonlayer}{nodelayer}
		\node [style=none] (0) at (-2.5, 1.5) {};
		\node [style=dot] (1) at (-2.5, 1) {};
		\node [style=dot] (2) at (-3.25, -0.5) {};
		\node [style=none] (3) at (-3.5, -1) {};
		\node [style=dot] (4) at (-1.75, -0.5) {};
		\node [style=none] (5) at (-1.5, -1) {};
		\node [style=none] (6) at (2.5, 1.5) {};
		\node [style=dot] (7) at (2.5, 1) {};
		\node [style=dot] (8) at (1.75, -0.5) {};
		\node [style=none] (9) at (1.5, -1) {};
		\node [style=dot] (10) at (3.25, -0.5) {};
		\node [style=none] (11) at (3.5, -1) {};
		\node [style=none] (12) at (-3.5, 1.5) {(a)};
		\node [style=none] (13) at (1.5, 1.5) {(b)};
		\node [style=none] (14) at (-3.5, 0.25) {$k+p$};
		\node [style=none] (15) at (-3.75, -1) {$p$};
		\node [style=none] (16) at (-2.5, -0.75) {$k$};
		\node [style=none] (17) at (-1.5, 0.25) {$k+l$};
		\node [style=none] (18) at (-1.25, -1) {$l$};
		\node [style=none] (19) at (1.5, 0.25) {$k+p$};
		\node [style=none] (20) at (1.25, -1) {$p$};
		\node [style=none] (21) at (2.5, -0.75) {$k$};
		\node [style=none] (22) at (2.5, -0.25) {$m$};
		\node [style=none] (23) at (3.5, 0.25) {$k+l$};
		\node [style=none] (24) at (3.75, -1) {$l$};
	\end{pgfonlayer}
	\begin{pgfonlayer}{edgelayer}
		\draw [style=external edge] (0.center) to (1);
		\draw [style=external edge] (3.center) to (2);
		\draw [style=external edge] (5.center) to (4);
		\draw [style=external edge] (6.center) to (7);
		\draw [style=massless external edge] (9.center) to (8);
		\draw [style=massless external edge] (11.center) to (10);
		\draw [style=massless edge] (1) to (2);
		\draw [style=massless edge] (2) to (4);
		\draw [style=massless edge] (7) to (8);
		\draw [style=massless edge] (7) to (10);
		\draw [style=edge] (8) to (10);
		\draw [style=massless edge] (1) to (4);
	\end{pgfonlayer}
\end{tikzpicture}}
  \caption{One loop contributions to the Sudakov form factor: (a)~from Ref.~\cite{Becher:2018gno}, and (b)~from Ref.~\cite{Jantzen:2011nz}.}
  \label{fig:form_factor}
\end{figure}

Examples for one-loop three-point functions are provided in \texttt{formfactor1L\_ebr}
and shown in Fig.~\ref{fig:form_factor}. The two integrals are taken
from Refs.~\cite{Becher:2018gno} and~\cite{Jantzen:2011nz}, respectively.
Diagram (a)  has only massless internal lines, but off-shell legs, defined in \texttt{generate\_formfactor1L\_massless\_ebr.py},
diagram (b)  has one massive propagator, while $p$ and $l$ are light-like, the corresponding generation file is called \texttt{generate\_formfactor1L\_massive\_ebr.py}.
We define $q=l-p, Q^2=-q^2>0$, $L^2=-l^2, P^2=-p^2$ and expand diagram (a) in
$L^2\sim P^2\ll Q^2$ and diagram (b) in the small mass limit.

\medskip

We would like to calculate diagram (a) for the point $Q^2=100$, $L^2=0.5$, $P^2=0.5$. For optimal performance with \pysecdec{}, all
terms in the integrand should be of order one and therefore we should factorise out $Q^2$.
We will introduce a smallness parameter $t$ by writing $L^2\rightarrow t\,L^2_r Q^2$, $P^2\rightarrow t\,P^2_r Q^2$,
where we introduced dimensionless parameters $P^2_r$ and $L^2_r$ of order one, and a dimensionless parameter $t\ll 1$. We can factorise out $Q^2$ from the $\mathcal{F}$ polynomial by adding a replacement rule $Q^2=1$ to the loop integral and introducing an additional prefactor of $Q^2$ raised to the power of the exponent of the $\mathcal{F}$ polynomial.

Expanding in $t$, there are four region vectors, $(-1,-1,-2,1)$, $(-1,0,-1,1)$, $(0,-1,-1,1)$ and $(0,0,0,1)$.
The original parameter values correspond to the new parameter values $Q^2=100$, $t=0.005$, $L^2_r=1$, $P^2_r=1$ and the leading order result in the smallness parameter (\mintinline{python}{expansion_by_regions_order=0}) is
\begin{align}
I&=(-1.7 {\cdot} 10^{-18} \pm 3.4 {\cdot} 10^{-18}) \, \eps^{-2} + \nn\\
 &+(-4.9 {\cdot} 10^{-16} \pm 4.0 {\cdot} 10^{-16} + (-3.5 {\cdot} 10^{-17}\pm 2.5 {\cdot} 10^{-17} ) \, i) \, \eps^{-1} + \nn\\
 &+(-0.313620 \pm 2.7 {\cdot} 10^{-10}  + (2.8 {\cdot} 10^{-11} \pm 1.1 {\cdot} 10^{-9}) \, i) \, + \nn\\
 &+\mathcal{O}\left(\eps, t\right).
\end{align}
Note that this diagram is finite and therefore the pole coefficients are numerically zero. For this relatively easy example with a scale ratio of order $10^2$, factorising out $Q^2$ does not make a big difference in the convergence behaviour, but we have illustrated the approach that will be more useful for larger scale differences and more difficult integrals.

\medskip

For diagram (b) we will similarly introduce $m^2=t\,Q^2$, factorise out $Q^2$ and expand in $t$. This integral requires an extra regulator. Instead of using an additional regulator in the powers of the propagators, we will use the \mintinline{python}{add_monomial_regulator_power='n'} option of \mintinline{python}{loop_regions}, which adds a monomial factor for each integration variable, with an extra regulator named \mintinline{python}{'n'} times a constant in the exponent of each of those factors, such that all the introduced exponents add up to zero. This makes the code generation many times faster for this example compared to using a regulator in the powerlist, however the compilation or the integration time is not significantly changed. A similar speedup of the code generation would apply to the previous one-loop box example if we would use the alternative regulator approach there.

There are 3 regions with region vectors $(-1,0,-1,1)$, $(0,-1,-1,1)$ and $(0,0,0,1)$.
The result for the point $Q^2 = 100, t = 0.01$, corresponding to $m^2=1$,  calculated at leading order (\mintinline{python}{expansion_by_regions_order=0}) in the smallness parameter $t$, reads
\begin{align}
I&=(-7.0 {\cdot} 10^{-18} \pm 7.4 {\cdot} 10^{-18}) \, \eps^{-2} + \nn\\
 &+(-1.3 {\cdot} 10^{-16} \pm 9.1 {\cdot} 10^{-17} + (2.7 {\cdot} 10^{-17}\pm 3.3 {\cdot} 10^{-17} ) \, i) \, \eps^{-1} + \nn\\
 &+(-0.138937 \pm 3.6 {\cdot} 10^{-10} + (-2.6 {\cdot} 10^{-10} \pm 3.3 {\cdot} 10^{-10}) \, i) \, + \nn\\
 &+ \mathcal{O}\left(\eps, n, t\right).
\end{align}

\subsection{Two-loop bubble in the large-mass expansion} 

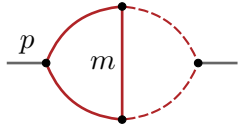
\begin{figure}[htb]
   \centering
   {\begin{tikzpicture}
	\begin{pgfonlayer}{nodelayer}
		\node [style=none] (0) at (-1.5, 0) {};
		\node [style=dot] (1) at (-1, 0) {};
		\node [style=dot] (2) at (1, 0) {};
		\node [style=none] (3) at (1.5, 0) {};
		\node [style=none] (4) at (-0.25, 0) {$m$};
		\node [style=none] (5) at (-1.25, 0.25) {$p$};
		\node [style=dot] (6) at (0, 0.75) {};
		\node [style=dot] (7) at (0, -0.75) {};
	\end{pgfonlayer}
	\begin{pgfonlayer}{edgelayer}
		\draw [style=external edge] (0.center) to (1);
		\draw [style=external edge] (2) to (3.center);
		\draw [style=edge, bend right] (1) to (7);
		\draw [style=edge, bend left] (1) to (6);
		\draw [style=edge] (6) to (7);
		\draw [style=massless edge, bend left] (6) to (2);
		\draw [style=massless edge, bend right] (7) to (2);
	\end{pgfonlayer}
\end{tikzpicture}}
   \caption{Two-loop bubble with three massive and two massless propagators.}
   \label{fig:bubble2L}
 \end{figure}

A two-loop two-point function example is provided in \texttt{bubble2L\_largem\_ebr}.
It performs the large-mass expansion of a bubble diagram with three massive and two massless propagators, called $\tilde{I}_{3}$ in Ref.~\cite{Fleischer:1996ju} and shown in Fig.~\ref{fig:bubble2L}.
  
\medskip

In the considered limit, the diagram has two regions with region vectors $(0,0,0,-1,-1,1)$ and $(0,0,0,0,0,1)$. The result for the point $p^2 = 0.002, m^{2} = 4$ up to leading order (\texttt{expansion\_by\_regions\_order=0}) in the smallness parameter $p^2$ reads
\begin{align}
I&=(-8.2 {\cdot} 10^{-11} \pm 1.0 {\cdot} 10^{-10} + (2.8 {\cdot} 10^{-11} \pm 8.9 {\cdot} 10^{-11} ) \, i) \, \eps^{-1} + \nn\\
 &+(+1.325116 \pm 2.3 {\cdot} 10^{-6} + (+3.92702 {\cdot} 10^{-1} \pm 2.1 {\cdot} 10^{-6}) \,i) \, + \nn\\
 &+\mathcal{O}\left(\eps, p^2/m^2\right).
\end{align}

\subsection{Two-loop bubble in the small-mass limit} 

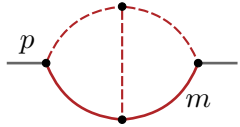
\begin{figure}[htb]
   \centering
    {\begin{tikzpicture}
	\begin{pgfonlayer}{nodelayer}
		\node [style=none] (0) at (-1.5, 0) {};
		\node [style=dot] (1) at (-1, 0) {};
		\node [style=dot] (2) at (1, 0) {};
		\node [style=none] (3) at (1.5, 0) {};
		\node [style=none] (4) at (1, -0.5) {$m$};
		\node [style=dot] (6) at (0, 0.75) {};
		\node [style=dot] (7) at (0, -0.75) {};
		\node [style=none] (8) at (-1.25, 0.25) {$p$};
	\end{pgfonlayer}
	\begin{pgfonlayer}{edgelayer}
		\draw [style=external edge] (0.center) to (1);
		\draw [style=external edge] (2) to (3.center);
		\draw [style=edge, bend right] (1) to (7);
		\draw [style=massless edge, bend left] (1) to (6);
		\draw [style=massless edge] (6) to (7);
		\draw [style=massless edge, bend left] (6) to (2);
		\draw [style=edge, bend right] (7) to (2);
	\end{pgfonlayer}
\end{tikzpicture}}
    \caption{Two-loop bubble with five propagators, two of them massive.}
   \label{fig:bubble2L_smallm}
 \end{figure}

This example is provided in \texttt{bubble2L\_smallm\_ebr} and performs the small-mass expansion.

\medskip

In the given limit, the diagram, shown in Fig.~\ref{fig:bubble2L_smallm}, has 4 regions with region vectors $(0,0,0,0,1)$, $(-1,-1,0,-1,0,1)$, $(-1,0,0,0,0,1)$, and $(0,-1,0,0,0,1)$. The result for the point $p^2 = 4, m^{2} = 0.002$, which starts at order 0 and is calculated up to order 1 in the smallness parameter $m^2$, reads
\begin{align}
I&=(-7.9 {\cdot} 10^{-10} \pm 1.5 {\cdot} 10^{-9} + (9.8 {\cdot} 10^{-10} \pm 1.5 {\cdot} 10^{-9} ) \, i) \, \eps^{-1} + \nn\\
 &+(+1.81362 \pm 1.5 {\cdot} 10^{-6} + (-7.541 {\cdot} 10^{-3} \pm 1.6 {\cdot} 10^{-6}) \, i) \, +\nn\\
 &+\mathcal{O}\left(\eps,\left(m^2/p^2\right)^2\right).
\end{align}

\subsection{Two-loop triangle}

\begin{figure}[htb]
   \centering
   {\begin{tikzpicture}
	\begin{pgfonlayer}{nodelayer}
		\node [style=dot] (0) at (-1.75, 0) {};
		\node [style=dot] (1) at (-0.75, 0.5) {};
		\node [style=dot] (2) at (0.25, 1) {};
		\node [style=dot] (4) at (-0.75, -0.5) {};
		\node [style=dot] (5) at (0.25, -1) {};
		\node [style=none] (6) at (0.75, -1) {};
		\node [style=none] (7) at (-2.25, 0) {};
		\node [style=none] (8) at (0.75, 1) {};
		\node [style=none] (9) at (-0.5, 0) {$m$};
		\node [style=none] (10) at (-2, 0.25) {$s$};
	\end{pgfonlayer}
	\begin{pgfonlayer}{edgelayer}
		\draw [style=external edge] (7.center) to (0);
		\draw [style=massless external edge] (5) to (6.center);
		\draw [style=massless external edge] (2) to (8.center);
		\draw [style=massless edge] (0) to (1);
		\draw [style=massless edge] (1) to (2);
		\draw [style=massless edge] (2) to (5);
		\draw [style=massless edge] (5) to (4);
		\draw [style=massless edge] (4) to (0);
		\draw [style=edge] (1) to (4);
	\end{pgfonlayer}
\end{tikzpicture}}
   \caption{Two-loop triangle with one massive propagator.}
   \label{fig:case8}
 \end{figure}
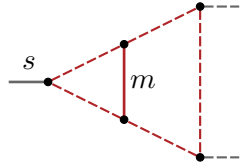

An example for a two-loop three-point function is given in \texttt{triangle2L\_ebr}.
It performs the large-mass expansion of a triangle diagram with one
massive propagator, calculated in Refs.~\cite{Fleischer:1997bq,Aglietti:2003yc}
 and shown in Fig.~\ref{fig:case8}.

\medskip

In the given limit, the diagram has 4 regions with region vectors
$(-1,-1,-1,-1,-1,0,1)$, $(-1,-1,0,0,0,0,1)$, $(0,0,-1,-1,-1,0,1)$, and
$(0,0,0,0,0,0,1)$. The result for the point $s = 0.002, m^{2} = 4$, which is calculated at leading order in the smallness parameter $s$ (\texttt{expansion\_by\_regions\_order=-1}), reads
\begin{align}
I&=(-1075.117 \pm 0.005 + (-392.705 \pm 0.003) \, i) \, \eps^{-2} + \nn\\
 &+(-6870.70 \pm 0.041 + (-8258.56 \pm 0.039) \, i) \, \eps^{-1} + \nn\\
 &+(-9444.07 \pm 0.363 + (-52842.92 \pm 0.374) \, i) \, + \nn\\
 &+\mathcal{O}\left(\eps,\left(s/m^2\right)^0\right).
\end{align}

\subsection{4-photon one-loop amplitude}
\label{sec:yyyy1L}

This example, contained in \texttt{yyyy1L}, calculates the one-loop 4-photon amplitude 
${\cal M}^{++--}$ to demonstrate the usage of \pysecdec{} for
amplitudes composed of linear combinations of master integrals and
coefficients which are rational functions of the kinematic invariants
and the regulators, i.e. $\eps=(4-D)/2$ in dimensional regularisation.

The amplitude for 4-photon scattering via a massless fermion loop can be expressed in terms of three independent helicity amplitudes, 
${\cal M}^{++++}$, ${\cal M}^{+++-}$, ${\cal M}^{++--}$, out of which the 
remaining helicity amplitudes forming the full amplitude can be reconstructed using crossing symmetry, Bose-symmetry and parity. Omitting an overall factor of $\alpha^2$ and a phase, the analytic expressions read (see e.g.~\cite{Karplus:1950zz,Bern:2001dg})
\begin{align}
{\cal M}^{++++} &= 8 \quad , \quad 
{\cal M}^{+++-} = -8 \;,\nn \\
{\cal M}^{++--} &= -8 \Bigl[  1 + \frac{t-u}{s} \log\left(\frac{t}{u}\right) 
 + \frac{t^2+u^2}{2 s^2} \Bigl( \log\left(\frac{t}{u}\right)^2 + \pi^2 \Bigr)\Bigr] \;\; .
\end{align}
%Up to an overall phase factor, $\frac{\langle 12 \rangle}{[12]}\frac{[34]}{\langle 34 \rangle}$, 
The amplitude ${\cal M}^{++--}$ can be expressed in terms of one-loop
4-point and 2-point integrals as
\begin{align}
{\cal M}^{++--} =-8 
 \left\{ 3(4-D) I_{4}^{D+4}(t,u) + \frac{t^{2}+u^{2}}{s} I_{4}^{D+2}(t,u) +  \frac{t-u}{s}\left(
    I_{2}^{D}(u)-I_{2}^{D}(t) \right) \right\}\,.
\label{eq:masters}
\end{align}
Note that $I_{4}^{D+4}$ is UV divergent, $I_{4}^{D+4}=1/(6\eps)+$finite,  and therefore provides the
rational part.

The file \texttt{generate\_yyyy1L.py} imports the integral definitions, contained in \texttt{integrals.py},
as well as the coefficients, contained in \texttt{coefficients.py}.
The integrals are given as a list containing the ``master'' integrals $I_{2}^{D}(u),I_{2}^{D}(t),$\\
$I_{4}^{D+2}(t,u),I_{4}^{D+4}(t,u)$.
The file \texttt{coefficients.py} contains the coefficient functions of these integrals, which need to be given as a list with the same ordering as the integral definitions.
Each coefficient function has the arguments \texttt{numerators, denominators, parameters}, where
the field \texttt{parameters} contains the names of the kinematic invariants. The polynomials in the numerator and denominator can also depend on the regulator $\eps$.
Note that the coefficient functions are provided as a list of lists, such that the same integral list can be used while iterating over different sets of coefficients.
This is relevant for example when calculating several helicity amplitudes depending on the same set of integrals.

For the kinematic point $t=-1.3,u=-0.8, s=-t-u$, the result reads
\begin{align}
  {\cal M}^{++--} &= (+0. \pm 2.1\cdot 10^{-16} )\,\eps^{-1} +  \nn\\
                  &+ (-28.431595834\pm 5.4\cdot 10^{-10}+
                          (-1.3\cdot 10^{-10} \pm 6.4\cdot 10^{-10}) \, i) \, + \nn\\
&+\mathcal{O}(\eps)\;.
\end{align}

\subsection{Example of a polynomial not related to Feynman integrals}

The example  \texttt{make\_regions\_ebr}  provides the expansion in the smallness parameter $t$ for the example polynomial given in \eqn{eq:examplepoly} multiplied by a monomial $x^\delta$:
\be
I=\int_0^\infty dx\, x^\delta (t+x+x^2)^{-1}\;.
\ee
The corresponding region vectors are $(0,1)$ and $(1,1)$.
Expanding in $t$ up to \texttt{expansion\_by\_regions\_order=0} and to order $\delta^0$ leads, for $t=0.01$, to
$I=4.60517\pm 4\cdot 10^{-16} + \mathcal{O}\left(\delta,t\right)$
in agreement with the analytic result $I=-\log(t)+{\cal O}(\delta,t)$.

%\clearpage

\section{Conclusions}
\label{sec:conclusion}
In this paper we have described the geometric formulation of the method of expansion by regions and its implementation within \pysecdec{}.
For integrals with a sufficiently large scale hierarchy, the time to reach a result with a given accuracy using expansion by regions stays approximately constant as the ratio of scales increases (i.e. the ``smallness parameter'' decreases), while the standard numerical evaluation by \pysecdec{} faces convergence problems, leading to a rapid increase of integration times and to uncertainties which by far exceed the uncertainties due to the truncation of the series in the ``smallness parameter''.

The new release also contains other major new features, the most important ones being an automated tuning of the integration contour and the ability to integrate weighted sums of integrals in a way which is optimised to reach a given accuracy goal on the sums rather than on the individual integrals.
The new features allow the efficient evaluation of multi-loop amplitudes in a largely automated way.

\section*{Acknowledgements}
We would like to thank Chaitanya Paranjape for very useful tests of
the development version of the code.
This research was supported in part by the COST Action CA16201 (``Particleface'') of the European Union
and by  the  Deutsche  Forschungsgemeinschaft (DFG, German Research Foundation) under grant 396021762 - TRR 257.
MK acknowledges supported by the Swiss National Science Foundation (SNF) under grant number 200020-175595.
SJ is supported by a Royal Society University Research Fellowship (Grant URF/R1/201268).

\renewcommand \thesection{\Alph{section}}
\appendix
\setcounter{section}{0}
\setcounter{equation}{0}
\section{Proof of the Cheng-Wu Theorem}
\label{sec:appendixWu}

To prove the Cheng-Wu theorem (an alternative proof is given in~\cite{Jantzen:2012mw}), we will show that we always get the same expression for the integrand after primary sector decomposition
(first defined in~\cite{Binoth:2000ps}),  independent of the choice of the coefficients $a_i$ in the expression
$\delta\left(1-  \sum_{i} a_ix_i\right)$.
In primary sector decomposition we divide the integration domain of an integral $I$ over $n$ parameters, constrained by a $\delta$-distribution of the above type,  into $n$ regions, where in the $j$-th region, $x_j$ is the largest integration variable:
\begin{align}
  I&=\int_0^\infty d{\bf x}\,f({\bf x})\delta\left(1-  \sum_{i} a_ix_i\right)\nn\\
    & =\sum_{j=1}^n\int_0^\infty d{\bf x}\left(\prod_{i\not =j}\theta(x_j-x_i)\right)f({\bf x})\,\delta\left(1-  \sum_{i} a_ix_i\right)\nn\\
  &=: \sum_{j=1}^n I_j\;,
\end{align}
where $d{\bf x}=\prod_{k}d x_k$ and the integrand $f({\bf x})$ must have the homogeneity property $f(\{\eta x_i\}) = \eta^{-N} f(\{x_i\})$.
Then, in each integral $I_j$ we substitute $x_i \rightarrow x_j t_i$ for $i\ne j$, leading to
\begin{align}
  I_j=\int_0^\infty dx_j\left( \prod_{i\not =j}\int_0^1 dt_i\right) x_j^{N-1}f(x_j,\{x_jt_i\}_{i\ne j})\,\delta\left(1-x_j(a_j+  \sum_{i\ne j} a_it_i)\right)\;.
\end{align}
Using the homogeneity property of $f$ we have $f(x_j,\{x_jt_i\}_{i\ne j})=x_j^{-N}f(1,\{t_i\}_{i\ne j})$, such that
\begin{align}
  I_j&=\int_0^\infty dx_j\,x_j^{-1}\left(\prod_{i\not =j}\int_0^1 dt_i\right) f(1,\{t_i\}_{i\ne j})\,\delta\left(1-x_jA\right)\;,\\
  A&=a_j+  \sum_{i\ne j} a_it_i\;.
\end{align}
Using $\int_0^\infty \frac{dx}{x}\,\delta\left(1-xA\right)=1$ we thus obtain
%Changing variables once more $x_jA\rightarrow t_j$ we obtain
\begin{align}
  I_j&= \left(\prod_{i\not =j}\int_0^1 dt_i\right) f(1,\{t_i\}_{i\ne j})\;,
\end{align}
which no longer depends on any of the $a_i$. 
Similarly, all the other primary sectors will also be independent of the $a_i$.

\section{Details about the two expansion methods}
\label{sec:appendix1}

\subsection{Convergence behaviour of the two methods}
\label{subsec:convergence_two_methods}

In Section~\ref{sec:ebralg}, we described two alternative methods to perform the expansion, one given by \eqn{eq:z-method}, which we will call the $z$-method, the other one given by \eqn{eq:t-method}, which we will call the $t$-method.
The user can employ either of these methods in \pysecdec, as described in Section \ref{sec:usage}.
The $t$-method usually leads to better convergence, because by construction the rescaling is such that no large scale differences remain in the integrand after the expansion with this method.
However, with the $t$-method, configurations can occur for special kinematic points which lead to a non-converging integrand because singularities occur at the upper border of the integration domain of some sectors.
The sector decomposition algorithm as implemented in \pysecdec{} relies on the fact that
(a) all endpoint singularities are located at the origin of Feynman parameter space (as given by the Landau equations before any remapping is performed), (b) integrable singularities are regulated by contour deformation, however the integration contour is fixed at the endpoints (zero and one in Feynman parameter space).
Therefore, singularities occurring at some Feynman para\-meter $x_i=1$ will not be regulated by contour deformation.
With the $t$-method, it can happen that the integrand vanishes at some $x_i=1$, which will lead to non-converging integrals.
Note that the original integral is usually well-defined, but the
combination of sector decomposition and contour deformation can
introduce spurious singularities.
How this can arise is explained below, using the simple example of a massive one-loop bubble integral in the small mass limit.

The second Symanzik polynomial $\mathcal{F}$ reads:
\begin{equation}
\mathcal{F} = -s x_{1} x_{2} + m^{2} (x_{1} + x_{2})^{2}\;.
\end{equation}
When performing the expansion by regions in the small mass limit, three regions are identified (one hard and two symmetric soft regions).
Let us focus on the soft region given by $\vec{v} = (-1,0,1)$. With the $t$-method, this corresponds to the following change of variables
\begin{align}
&x_{1} \rightarrow x_{1}/m^{2} \nn\\
&x_{2} \rightarrow x_{2} \nn\\
&m^{2} \rightarrow m^{2} \;,
\end{align}
such that $\mathcal{F}$ can be written as:
\begin{equation}
\mathcal{F} = -\frac{s}{m^{2}} x_{1} x_{2} + m^{2} \left( \frac{x_{1}}{m^{2}} + x_{2}\right)^{2} \;.
\end{equation}
Keeping the leading order terms for $m^{2} \rightarrow 0$ leads to
\begin{equation}
\mathcal{F} \approx -\frac{s}{m^{2}} x_{1} x_{2} +  \frac{x_{1}^{2}}{m^{2}}\;.
\end{equation}
The sector decomposition algorithm splits the region integral into two sectors.
Here we show the $\mathcal{F}$ polynomial of the first sector which is obtained by setting $x_1=1,~x_2=y$:
\begin{equation}
  \mathcal{F}_{1} = -\frac{s}{m^{2}} y +  \frac{1}{m^{2}}=\frac{1}{m^{2}}\left( -s\, y+1\right)\;.
  \label{eq:factored}
\end{equation}
When $s=1$ we therefore have $\mathcal{F}_{1} = 0$ at the upper border of the integration domain, $y=1$.  

Let us see now what happens instead in the $z$-approach. We write $\mathcal{F}$ as
\begin{equation}
\mathcal{F} = -s x_{1} x_{2} + z\,m^{2} (x_{1} + x_{2})^{2}\;,
\end{equation}
where we already changed $m^{2}$ to $z\,m^{2}$. The change of variables for the integration variables reads:
\begin{align}
&x_{1} \rightarrow x_{1}/z \nn\\
&x_{2} \rightarrow x_{2} \;,
\end{align}
leading to
\begin{equation}
\mathcal{F} = -\frac{s}{z} x_{1} x_{2} + z\,m^{2} \left( \frac{x_{1}}{z} + x_{2}\right)^{2} \;.
\end{equation}
Keeping the leading order terms in the expansion in $z$:
\begin{equation}
\mathcal{F} \approx -\frac{s}{z} x_{1} x_{2} +  m^{2} \frac{x_{1}^{2}}{z}\;.
\end{equation}
Setting $z=1$ and looking at $\mathcal{F}_{1}$ as before we therefore have
\begin{equation}
\mathcal{F}_{1} = -s \,y +  m^{2}\;.
\end{equation}
This function vanishes at $y = 1$ when $s = m^{2}$, which by definition is not the case since we are expanding in the small mass limit.
Therefore the potential problem at the upper boundary is absent with the $z$-method, however we see that large scale differences remain in the integrand.

Another option which avoids convergence issues is to use $t=\frac{m^2}{2s}$ as expansion parameter.
After factoring out overall powers of $s$ and $t$ one obtains for the soft region in the limit $t\to 0$:
\begin{equation}
\mathcal{F} = \frac{s}{t}\, x_1 \left( -x_2 + 2 x_1\right) + \mathcal{O}(t^0).
\end{equation}
For this region sector decomposition leads to two sectors with graph polynomials (omitting overall factors)
\begin{equation}
\mathcal{F}_1  = -y+2 \quad\mathrm{and}\quad \mathcal{F}_2 = y \left( -1 + 2 y \right). 
\end{equation}
Here no large scale differences remain in the integral and integrable singularities are kept away from the integration boundaries at $y=0$ and $y=1$.

\subsection{Proof that the two alternative expansion methods are equivalent}

We will show here that the $z$-method and the $t$-method described in Section~\ref{sec:ebralg} are equivalent.
To this aim we introduce the operators
\begin{align*}
W_t: (x_i, t)&\to (t^{v_i} x_i,t), &V_z: (x_i,t) &\to (z^{v_i} x_i , z t), \\
W_{t^{-1}}:(x_i,t)&\to (t^{-v_i} x_i, t), &R_{z\to1}: z &\to 1,
\end{align*}
where the substitutions $V_z$ and $W_t$ were previously defined in~\eqref{eq:z-method} and~\eqref{eq:t-method} respectively.
The operator $T_y = \sum_n T_y^n$ performs a power series expansion in the small parameter $y$, and $T_y^n$ projects out the order $n$ term of the series.

Using these definitions we can describe the two expansion methods applied to an integrand $f(\{x_i\},t)$ as:
\begin{align}
\text{$z$-method: }\; & f_z(\{x_i\},t)=\mkern-9mu& \sum_n f^n_z(\{x_i\},t) &= \sum_n R_{z\to 1}T^n_z V_z f(\{x_i\},t) \\
\text{$t$-method: }\; & f_t(\{x_i\},t)=\mkern-9mu& \sum_n f^n_t(\{x_i\},t)  &= \sum_n T^n_t W_t f(\{x_i\},t).
\end{align}
We will show that the integrand expansion terms are related by a rescaling of the integration parameters $x_i$ as
\begin{equation}
f^n_t (\{x_i\},t)= W_t f^n_z(\{x_i\},t).
\label{eq:tzrelation}
\end{equation}
Noting that $W_{t^{-1}} W_t = \mathbb{I}$ and $R_{z\to 1} V_z = \mathbb{I}$, we insert identity operators into the definition of the expansion terms:
\begin{alignat}{3}
 f^n_z(\{x_i\},t)= {}&&{} \mathbb{I} R_{z\to 1} T^n_z V_z f(\{x_i\},t) ={}&&{} W_{t^{-1}} W_t R_{z\to  1}  T^n_z V_z f(\{x_i\},t)\;,\nn\\
 f^n_t(\{x_i\},t)={}&&{} T^n_t W_t \mathbb{I} f(\{x_i\},t) ={}&&{}  T^n_t W_t R_{z\to  1} V_z  f(\{x_i\},t)\;. \label{eq:altways_met1}
\end{alignat}
In  $f^n_z$ the rescaling $W_t$ can be moved past $R_{z\to 1}T^n_z$ since the operations commute.
Similarly $R_{z\to 1}$ can be moved to the left in $f^n_t$ which results in
\begin{alignat}{2}
 f^n_z(\{x_i\},t)={}&&{}  W_{t^{-1}} R_{z\rightarrow 1} T^n_z W_t V_z f(\{x_i\},t)\;,\nn\\
 f^n_t(\{x_i\},t)={}&&{}  R_{z\rightarrow 1} T^n_t W_t V_z f(\{x_i\},t)\;.
\end{alignat}
Apart from the expected factor $W_{t^{-1}}$ in $f^n_z$, the above equations differ only by the change of $T^n_t\rightarrow T^n_z$.
We will show that in this case $T^n_t$ and $T^n_z$ are actually the same. To see this, we note that $W_t V_z f(\{x_i\},t) = f(\{(zt)^{v_i}x_i\},zt)$. The dependence on $t$ and $z$ is always of the form $(zt)$. For any function $g(zt)$, we could do a series expansion in terms of $(zt)$ and it would automatically already be a series expansion in $t$ and simultaneously a series expansion in $z$:
\begin{equation}
T_{(zt)} g(zt) = T_z g(zt) = T_t g(zt).
\end{equation}
Since this relation is also valid term by term in the expansion, we have shown that \eqref{eq:tzrelation} is true.
The derived relation for the integrand terms implies the equivalence of the corresponding integrals up to a change of variables:
\begin{equation}
\int_{\mathbb{R}^N_{\geq 0}} \frac{\mathrm{d}\mathbf{x}}{\mathbf{x}} f^n_z(\mathbf{x},t) 
= \int_{\mathbb{R}^N_{\geq 0}} \frac{\mathrm{d}\mathbf{x}}{\mathbf{x}} W_t f^n_z(\mathbf{x},t)
= \int_{\mathbb{R}^N_{\geq 0}} \frac{\mathrm{d}\mathbf{x}}{\mathbf{x}} f^n_t(\mathbf{x},t).
\end{equation}
In the first step the integration variables are rescaled by applying $W_t$, while \eqref{eq:tzrelation} is used in the second step.

%\section*{References}
\addcontentsline{toc}{chapter}{References}

\bibliographystyle{JHEP}

\renewcommand*{\bibfont}{\justify}
\bibliography{main_EbR}

\end{document}